\documentclass[pra,aps,twocolumn,nopacs,superscriptaddress,nofootinbib]{revtex4}
\usepackage[bbgreekl]{mathbbol}
\usepackage{appendix}
\usepackage[T1]{fontenc}
\usepackage[latin1]{inputenc}
\usepackage{lmodern}
\usepackage{graphicx}
\usepackage{dcolumn}
\usepackage{bm}
\usepackage{amsmath}
\usepackage{amssymb}
\usepackage{pst-all}
\usepackage{psfrag}
\usepackage{epsfig}
\usepackage{titletoc}
\usepackage{minitoc}

\def\ii{{\rm i}} \def\ee{{\rm e}}  
\def\Eb{{\bf E}} \def\Rb{{\bf R}} \def\rb{{\bf r}}  
  
 \def\pb{{\bf p}}

\def\zz{\hat{\bf z}}

\begin{document}

\title{Electrically tunable nonlinear plasmonics in graphene nanoislands}

\author{Joel D. Cox}
\affiliation{ICFO-Institut de Ciencies Fotoniques, Mediterranean Technology Park, 08860 Castelldefels (Barcelona), Spain}
\author{F. Javier Garc\'{\i}a de Abajo}
\email{javier.garciadeabajo@icfo.es}
\affiliation{ICFO-Institut de Ciencies Fotoniques, Mediterranean Technology Park, 08860 Castelldefels (Barcelona), Spain}
\affiliation{ICREA-Instituci\'o Catalana de Recerca i Estudis Avan\c{c}ats Barcelona, Spain}

\begin{abstract}
Nonlinear optical processes rely on the intrinsically weak interactions between photons enabled by their coupling with matter. Unfortunately, many applications in nonlinear optics are severely hindered by the small response of conventional materials. Metallic nanostructures partially alleviate this situation, as the large light enhancement associated with their localized plasmons amplifies their nonlinear response to record high levels. Graphene hosts long-lived, electrically tunable plasmons that also interact strongly with light. Here we show that the nonlinear polarizabilities of graphene nanoislands can be electrically tuned to surpass by several orders of magnitude those of metal nanoparticles of similar size. This extraordinary behavior extends over the visible and near-infrared for islands consisting of hundreds of carbon atoms doped with moderate carrier densities. Our quantum-mechanical simulations of the plasmon-enhanced optical response of nanographene reveal this material as an ideal platform for the development of electrically tunable nonlinear optical nanodevices.
\end{abstract}
\maketitle
\tableofcontents

% {\bf KEYWORDS:} graphene, plasmons, nonlinear response, optical tunability

\section{Introduction}

The well-established field of nonlinear photonics hosts a vast number of applications, including spatial and spectral control of laser light, all-optical signal processing, ultrafast switching, and sensing \cite{B08_3, G13}. Because the efficiencies of nonlinear optical processes are generally poor, considerable effort has been devoted towards seeking materials that can display nonlinear effects at low light intensities and ultrafast response times \cite{KZ12, G13, KS13}. For this purpose, plasmonic nanostructures have been particularly attractive due to their ability to generate high local intensity enhancements through strong confinement of electromagnetic fields \cite{S11, KZ12, AP10}, leading to second-harmonic polarizabilities as high as $\sim10^{-27}$ esu per atom, as measured for noble metal nanoparticles \cite{KZ12}, and even beating the best molecular chromophores \cite{VLH98,RBB07,KZ12}. However, although localized plasmons can be customized through the size, shape, and surrounding environment of the metal nanostructures \cite{S11}, they suffer from low lifetimes and lack post-fabrication tunability \cite{BL12}.

Doped graphene has recently attracted much attention as an alternative plasmonic material capable of sustaining electrically tunable optical excitations with long lifetimes \cite{paper176, BL12, GPN12, paper235, JGH11, YLL12, FAB11, YLC12, paper196, FRA12}. The existence of gate-tunable plasmons in graphene has been confirmed by THz \cite{JGH11, YLL12} and mid-infrared \cite{FAB11, YLC12} spectroscopies, while optical near-field microscopy has been used to image them in real space \cite{paper196, FRA12}. Efforts to extend the plasmonic response of graphene to the visible and near-infrared regimes are currently underway \cite{paper235}. Additionally, graphene has been predicted to display intense nonlinearity due to its anharmonic charge-carrier dispersion relation \cite{M07_2}. Recent four-wave mixing \cite{HHM10}, Kerr effect \cite{ZVB12}, and third-harmonic generation \cite{KKG13} experiments already confirm a large third-order response in this material in the undoped, plasmon-free state. Graphene plasmons could amplify this response further \cite{M11}, and even enable strong few-photon interactions in small islands \cite{paper226,MSG14}.

\begin{figure*}
\includegraphics[width=1\textwidth]{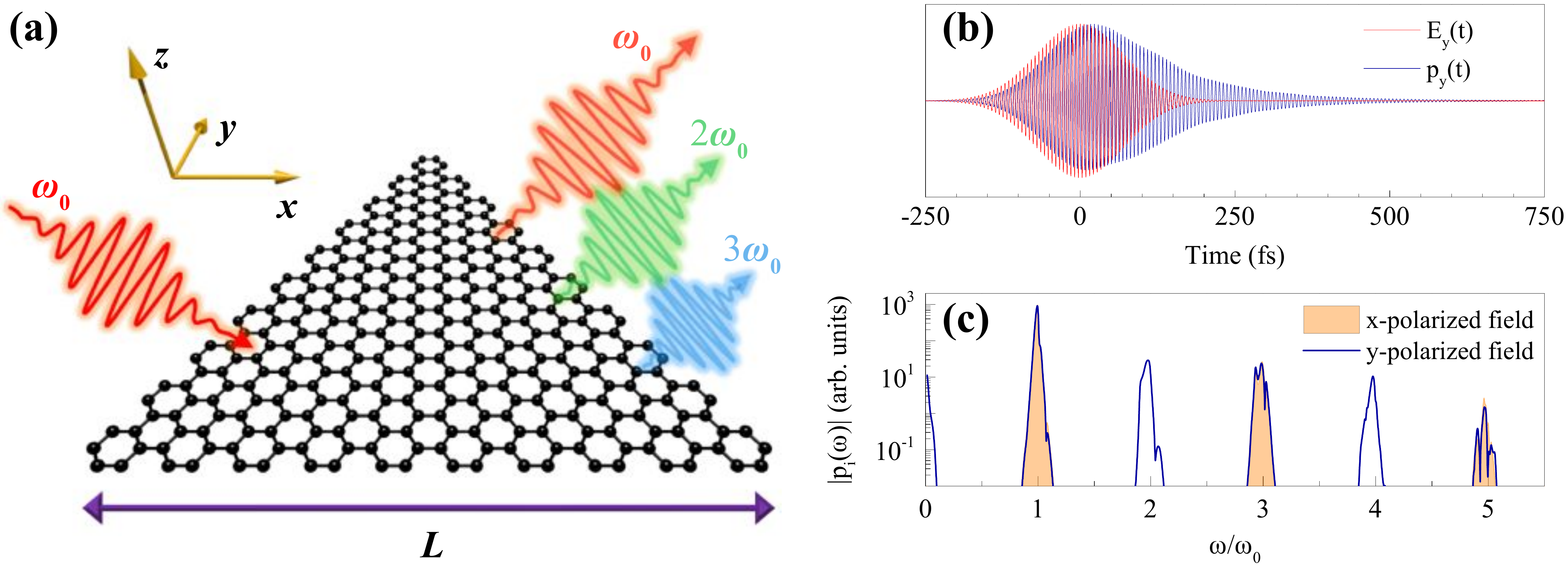}
\caption{{\bf Nonlinear response of nanographene.} {\bf (a)} Illustration of a doped graphene nanotriangle (armchair edges, $N=330$ carbon atoms, $L=4.1\,$nm side length, doped with $Q=2$ additional charge carriers) under irradiation by a short light pulse ($166\,$fs FWHM duration, $10^{12}$\,W$/$m$^2$ peak intensity, $\hbar\omega_0=0.68\,$eV central energy) tuned to one of the graphene plasmons. {\bf (b)} Time variations of the incident electric field and the induced graphene dipole. {\bf (c)} Harmonic analysis of the graphene dipole for polarizations along the $x$ and $y$ directions (see axes in (a)).}
\label{fig1}
\end{figure*}

Here we show that the nonlinear optical polarizabilities of small graphene nanoislands ($<10\,$nm) exceed by several orders of magnitude those of the best conventional nonlinear materials, including noble metal nanoparticles of similar lateral size (but obviously of much greater volume). Nonlocal and finite-size effects dominate the response of these structures \cite{paper183,paper215}, which we model in a quantum-mechanical fashion. Specifically, we perform density-matrix simulations using a tight-binding description of the electronic states and following complementary time-domain and perturbative approaches (see details in Methods and Appendix). Our results reveal unprecedented levels of nonlinearity when the graphene nanoislands are doped with only a few electrons.

\section{Results and discussion}

Fig.\ \ref{fig1} illustrates the optical nonlinearity of one of the armchair-edged triangular graphene nanoislands considered in this work, consisting of 330 carbon atoms, spanning a side length of $4.1\,$nm, and doped with two additional charge carriers ($2.3\times10^{13}\,$cm$^{-2}$ doping density, equivalent to 0.56\,eV Fermi energy in extended graphene). Upon illumination with a light pulse of central energy $\hbar\omega_0=0.68\,$eV, tuned to one of its plasmons, the island is capable of producing significant nonlinear polarization at multiple harmonics (Fig.\ \ref{fig1}a), including second- and third-harmonic generation (SHG and THG). The temporal evolution of the induced polarization (Fig.\ \ref{fig1}b, blue) and its spectral decomposition (Fourier transform, Fig.\ \ref{fig1}c) show the excitation of high harmonics using a relatively moderate pulse fluence $177\,$mJ$/$m$^2$ (see Fig.\ \ref{figS1} in the Appendix for additional results obtained from various incident fluences). Although graphene is a centrosymmetric 2D crystal, which ordinarily would prevent even-ordered nonlinear processes from occuring \cite{B08_3}, this symmetry can be broken by the finite size of the nanoisland. Notice that the mirror symmetry of the nanoisland along $x$ results in the vanishing of even-harmonic generation when the incident light is polarized along that direction. Conversely, both odd and even harmonics are observed with a pulse polarized along the asymmetric $y$ direction. Interestingly, the intensity of odd harmonics is independent of polarization direction because the nanoisland has threefold rotational symmetry. For the pulse polarized along the $y$-direction, we also take special note of a static electric field formed in the graphene nanoisland due to optical rectification, as indicated by the polarization emerging at zero frequency in Fig.\ \ref{fig1}c.

\begin{figure*}
\includegraphics[width=1\textwidth]{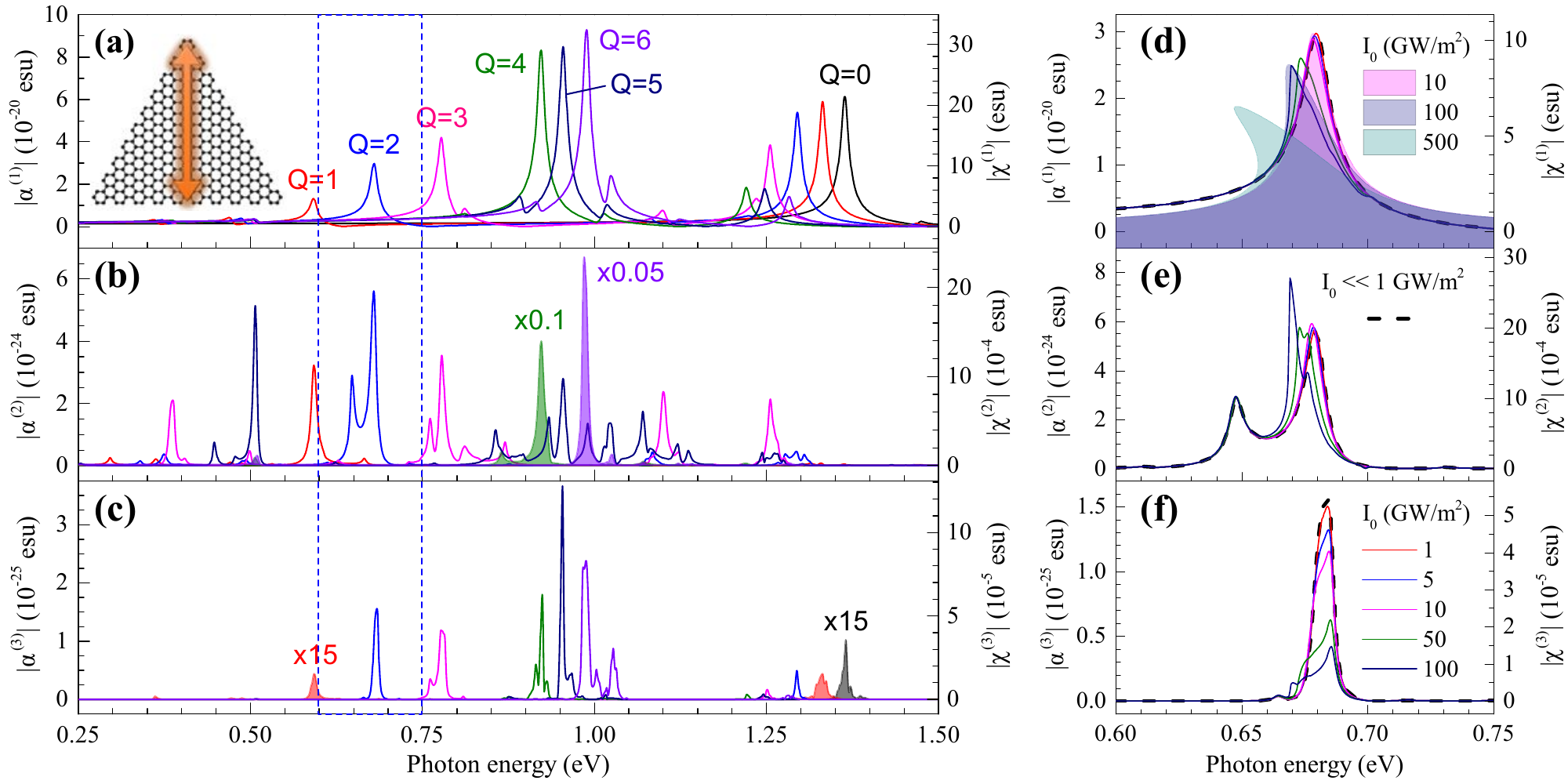}
\caption{{\bf Linear and nonlinear polarizability spectra.} We study the nanotriangle of Fig.\ \ref{fig1} for external polarization perpendicular to one of the graphene sides, assuming different doping levels as indicated by the number of additional charge carriers $Q$. {\bf (a-c)} Linear (a), second-harmonic (b), and third-harmonic (c) polarizabilities for low-intensity continuous illumination. {\bf (d-f)} Same as (a-c) calculated for various high intensities $I_0$ (see legends), and at frequencies near the lowest-energy dipole plasmon under $Q=2$ doping, as indicated by the dashed blue box spanning (a-c). The filled curves in (b) and (c) have been multiplied by the factors indicated with text of the corresponding color, while the filled curves in (d) are obtained with a classical anharmonic oscillator model (see main text).}
\label{fig2}
\end{figure*}

\begin{figure*}
\includegraphics[width=1\textwidth]{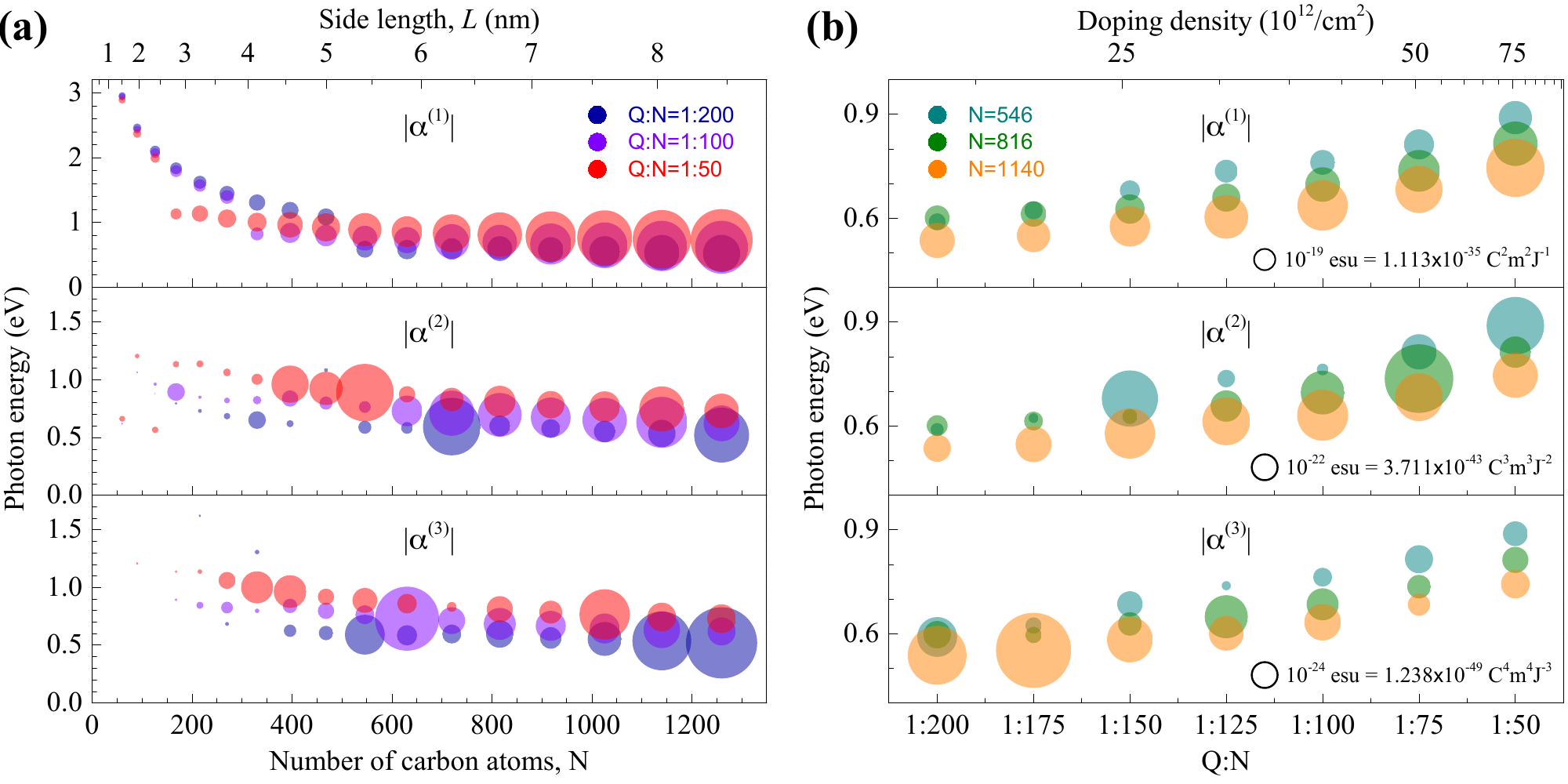}
\caption{{\bf Evolution of the nonlinear response with graphene size and doping level.} We show linear and nonlinear peak polarizabilities (proportional to symbol areas, see empty-circle scales in (b)) at the dominant plasmon features of graphene armchair nanotriangles as a function of their size (a) and doping level (b) for selected doping levels and sizes, respectively. The vertical positions of the symbols indicate the resonant incident light energies.}
\label{fig3}
\end{figure*}

\begin{figure*}
\includegraphics[width=0.8\textwidth]{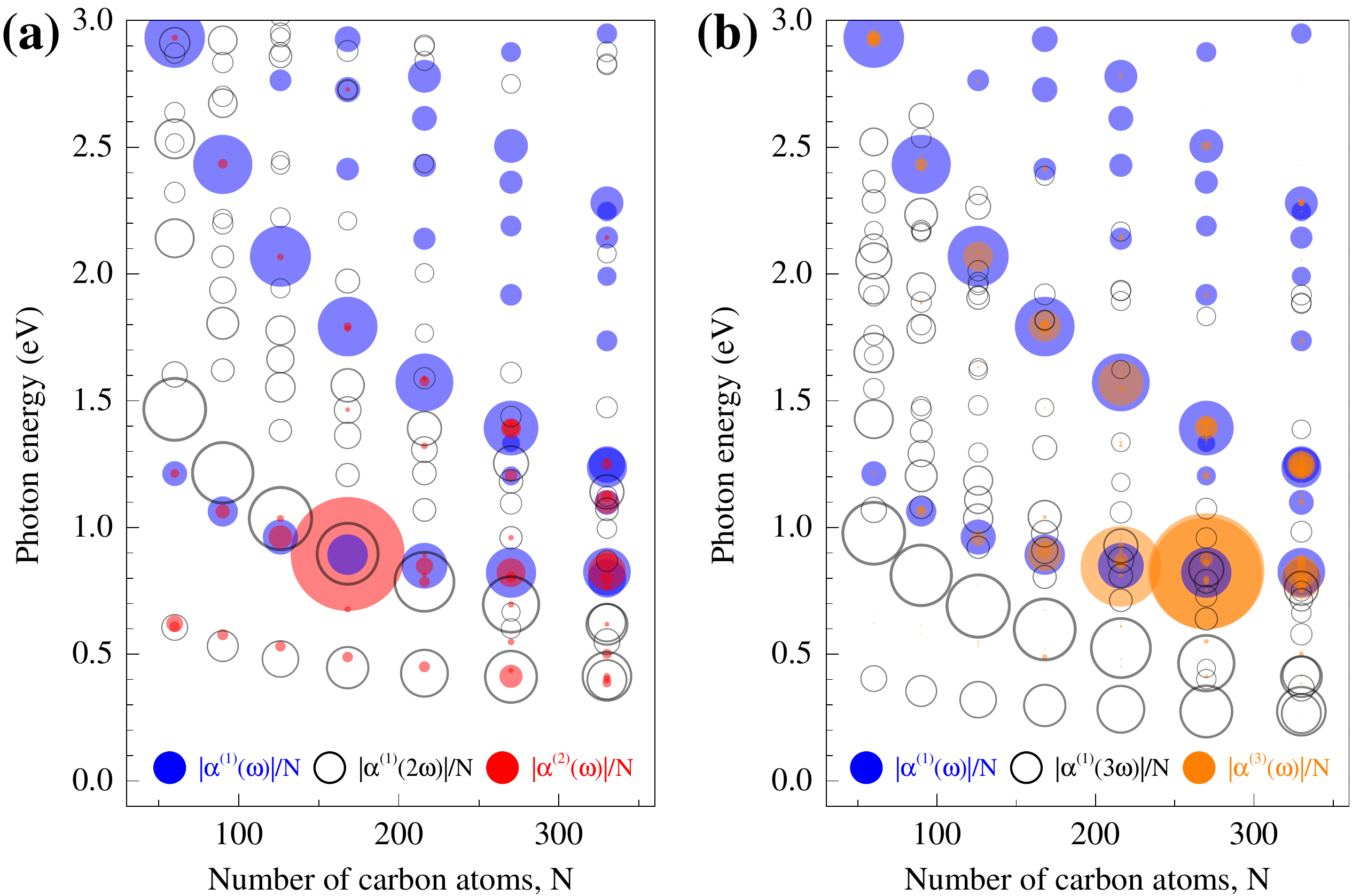}
\caption{{\bf Doubly resonant SHG and THG.} Local maxima in the polarizability spectra, normalized to the number of carbon atoms, are shown for various nanoislands with a fixed doping carrier density $N/Q=100$. The filled symbols show the linear and nonlinear polarizabilities as functions of the fundamental incident photon frequency, while the empty symbols show the linear polarizability at the indicated harmonic frequency (see legends). Large SHG (a) and THG (b) is observed for nanoislands that support plasmons at both the fundamental frequency and twice or three times that frequency, respectively. The polarizability is proportional to the area of the symbols.}
\label{fig4}
\end{figure*}

\begin{figure*}
\includegraphics[width=1\textwidth]{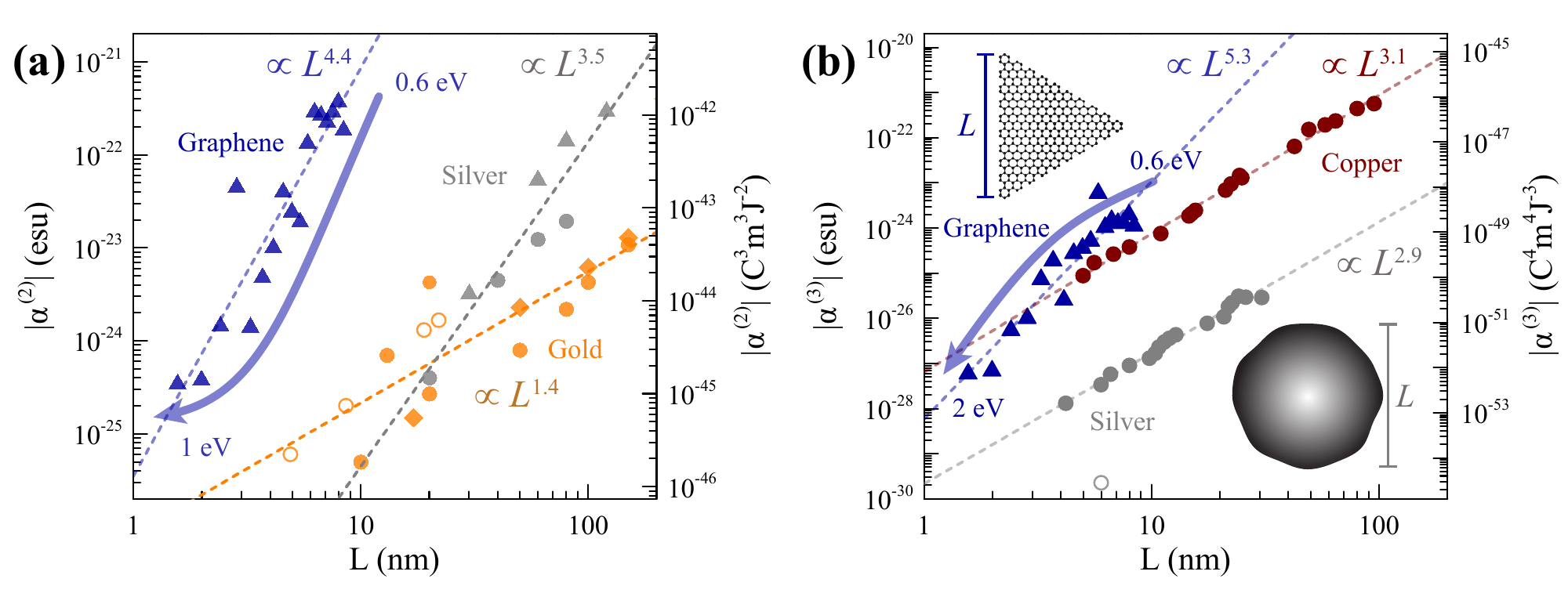}
\caption{{\bf Second- and third-harmonic responses of nanographene compared with noble metal nanoparticles.} We show the nonlinear polarizabilities of graphene nanotriangles as a function of their side length $L$ at a fixed doping of one electron per 100 carbon atoms, compared with measured values of noble metal nanoparticles (MNPs) of the same maximum length. In (a) we show second harmonic data for roughly-spherical gold and silver MNPs measured at fundamental wavelengths of 800 and $780\,$nm, respectively (solid circles) \cite{VLH98, RBB07}; roughly-spherical gold MNPs measured at a fundamental wavelength of $1064\,$nm (open circles) \cite{GBG99}; gold decahedra measured at a fundamental wavelength of $790\,$nm (solid diamonds) \cite{RDB10}; and silver triangles measured at a fundamental wavelength of $1064\,$nm (solid triangles) \cite{SSN09}. In (b) we show degenerate four-wave mixing data for silver and copper MNPs measured at wavelengths of $\sim420$ and $\sim570\,$ nm, respectively (solid circles) \cite{UKO94}, as well as THG data for silver MNPs measured at a fundamental wavelength of $1230\,$nm \cite{LTY06} (open circle at the bottom). Dashed lines are added to indicate the approximate scaling with $L$ according to experimental observations. The calculated graphene polarizabilities correspond to incident light energies as indicated along the curved arrows.}
\label{fig5}
\end{figure*}

For a quantitative analysis of the linear and nonlinear optical response of nanographene, we examine the incident-frequency dependence of the nonlinear polarizability in Fig.\ \ref{fig2}. Both time-domain and perturbative approaches produce nearly identical results when the former is computed for low light intensities (see Appendix). For simplicity, we concentrate on the same nanotriangle as in Fig.\ \ref{fig1}, under different doping conditions, ranging from $Q=0$ to $Q=6$ additional charge carriers. In undoped graphene ($Q=0$), the linear polarizability (Fig.\ \ref{fig2}a) is dominated by a single $>1\,$eV plasmon in the spectral region under consideration, which however produces only negligible SHG (Fig.\ \ref{fig2}b) and THG (Fig.\ \ref{fig2}c). In contrast, when the island is doped, $<1\,$eV plasmon features emerge (Fig.\ \ref{fig2}a), which move towards higher energies as $Q$ is increased. These highly-tunable, low-energy dipole plasmon modes are a result of the collective nature of the electronic excitations of the nanoisland, and do not coincide with any dominant electron-hole pair transitions \cite{paper214,paper235}. Incidentally, zigzag edges are detrimental for the emergence of these plasmons and the resulting tunable nonlinear response (see Figs.\ \ref{figS4} and \ref{figS5} in the Appendix). Importantly, intense features appear in the nonlinear spectra at incident photon energies tuned to the low-energy plasmons (Fig.\ \ref{fig2}b,c). These features exhibit multicomponent line profiles that indicate a complex interplay between the responses at the incident frequency and at multiples of that frequency. The nonlinear strengths follow a non-monotonic evolution with $Q$ that we also attribute to that interplay (see below). For quantitative comparisons with bulk materials, we approximate the nonlinear susceptibilities $\chi^{(2)}$ and $\chi^{(3)}$ in doped nanographene by considering the polarizability per atom (denoted here as $\tilde{\alpha}^{(s)}$ for the harmonic $s$) and calculating $\chi^{(s)}=\tilde{\alpha}^{(s)}n_C/d_{\rm gr}$, where $n_C=3.8\times10^{15}$\,cm$^{-2}$ is the areal density of carbon atoms in graphene and $d_{\rm gr}\simeq3.3\times10^{-8}$\,cm is the effective thickness of a graphene layer (see right vertical axes in Fig.\ \ref{fig2}). For comparison, we note that the third-order susceptibility of extended graphene without the involvement of any plasmons has been measured via four-wave mixing experiments as $|\chi^{(3)}|\sim10^{-7}$\,esu, exceeding the maximum value obtained for a 4\,nm thick gold film measured under the same experimental conditions \cite{HHM10}. Additionally, third-harmonic generation measurements have yielded values $|\chi^{(3)}|\sim10^{-8}$\,esu in graphene \cite{KKG13}.

The linear and nonlinear spectral lineshapes display a dependence on light intensity that is clearly observable above $\sim1\,$GW$/$m$^2$, as shown through time-domain simulations in Fig.\ \ref{fig2}d-f, obtained from the Fourier transform of the induced dipole over an optical cycle after reaching a steady-state regime (see Appendix). In particular, the third-order nonlinearity contributes to the polarization oscillating at the incident frequency $\omega$ through the Kerr effect \cite{B08_3}, the spectral details of which can be actually mimicked by a classical model, assuming that each electron of mass $m_e$ oscillates around its equilibrium position $x=0$ driven by the incident field and subjected to an anharmonic potential $U(x)=-(m_e\omega_0^2/2) x^2-(m_ea/4)x^4$ of resonance energy $\hbar\omega_0=0.68\,$eV and nonlinear coefficient $a$. Using a fixed value $a=-(4.7+0.3\,\ii)\times10^{47}\,$m$^{-2}$s$^{-2}$, this model reproduces the initial redshift with increasing intensity $I_0$ found in the calculated linear polarizability for $I_0=10^{10}-10^{11}\,$W$/$m$^2$ (see Fig.\ \ref{fig2}d and Appendix), while it predicts optical bistability at even higher intensities ($I_0=5\times10^{11}\,$W$/$m$^2$).

The number of electrons involved in the observed resonances, as estimated from the $f$-sum rule \cite{PN1966,paper235}, is roughly proportional to the resonance frequency times the maximum polarizability, where the latter is given by the area of the symbols in the top panel of Fig.\ \ref{fig3}a. Although the largest islands studied here involve many electrons in each resonance according to this criterium, and thus, we can legitimately talk about plasmon resonances, it is interesting to note that the mode examined in Fig.\ \ref{fig2}d for the $330$-atom triangle has an effective strength of only $\sim0.3$ electrons. However, the energy of this mode is very different from those of individual dipole-active electron-hole pair excitations \cite{paper214}, which emphasizes its many-body character. It is nonetheless surprising that nonlinear effects are observed at relatively moderate intensities (e.g., $10^{11}\,$W$/$m$^2$), for which the excited state still has a low population $n_p\sim0.08$, as estimated by equating the energy absorbed by the particle (i.e., $\sigma^{\rm abs}I_0$, where $\sigma^{\rm abs}\approx(4\pi\omega/c){\rm Im}\{\alpha^{(1)}\}$ is the absorption cross section) to the energy dissipated through plasmon decay (i.e., $n_p\hbar\omega_p/\tau$, where $\omega_p$ and $\tau$ are the mode frequency and decay time, respectively).

We present in Fig.\ \ref{fig3} an overview of the dependence of the maximum first-, second-, and third-harmonic polarizabilities on island size and doping (see also Fig.\ \ref{figS2} in the Appendix for the polarizabilities normalized to the number of atoms). The linear polarizabilities increase monotonically with size and doping, while the plasmon energies scale roughly as $\sim Q^{1/4}/N^{1/2}$, in a way that is consistent with a local classical description of the optical response \cite{paper235}. With fixed doping densities, the maximum linear polarizability is found among the higher-energy plasmon modes for smaller islands, as illustrated in Fig.\ \ref{fig2}a for low doping. As the nanoisland size increases, and with it the doping level, the lower-energy modes eventually become dominant. In contrast, the nonlinear polarizabilities exhibit a non-monotonic evolution with both $Q$ and $N$, which we again attribute to the presence of resonances at both the fundamental and the harmonic frequencies. When normalized per carbon atom, the THG susceptibilities take extraordinarily large values $|\chi^{(3)}|\sim10^{-6}$--$10^{-4}$\,esu (see Appendix and below).

A mechanism of double plasmonic enhancement is illustrated in Fig.\ \ref{fig4}, showing anomalously large nonlinear polarizabilities when a plasmon exists at a multiple of the fundamental frequency, which is in turn tuned to another plasmon. This phenomenon has been recently invoked to predict nonlinearities in graphene islands at the single-photon level \cite{MSG14}. For SHG, Fig.\ \ref{fig4}a shows this effect taking place for an island consisting of $N=168$ atoms. Similarly, Fig.\ \ref{fig4}b shows large THG with $N=270$. The non-monotonic evolution of the nonlinear polarizabilities with size and doping noted in Figs.\ \ref{fig2} and \ref{fig3} are also due to this type of double resonance effect. For a fixed geometry, the double enhancement phenomenon should be attainable by varying the doping charge density: consider, for example, the highly-tunable, low-energy dipole plasmon mode of Fig.\ \ref{fig2}a, which evolves with doping, eventually converging to the less-tunable, higher-energy mode, so that at some point before converging the two modes satisfy the required 1:2 energy ratio. This phenomenon should also occur in imperfect nanoislands with a predominance of armchair edges, although higher levels of doping could be necessary (see Fig.\ \ref{figS6} in the Appendix).

Fig.\ \ref{fig5}a shows that the SHG polarizabilities predicted for nanographene exceed by three orders of magnitude the optimal values measured for noble metal nanoparticles of similar lateral size. For a comprehensive comparison of SHG with noble metals, we present data from experiments performed on various nanoparticle morphologies, including highly asymmetric structures, as well as for different excitation frequencies \cite{RBB07, GBG99, RDB10, SSN09}. The comparison per unit volume is even more favorable to graphene, as it is an atomically thin structure, in contrast to the three dimensional nanoparticles. The third-order polarizability in graphene is also above that measured for copper and silver \cite{UKO94, LTY06} (see Fig.\ \ref{fig5}b), and more so when considering that these measurements refer to fully degenerate four-wave mixing experiments, which tend to yield significantly higher values than those obtained when looking at THG or non-degenerate four-wave mixing \cite{BSD14}. Although larger graphene islands present a computational challenge beyond our current computational capabilities, the pace at which their nonlinear polarizabilities increase with size should be faster than that of noble metals, as expected from the extrapolation of the dashed curves in Fig.\ \ref{fig5}. It should be noted that this occurs at plasmon energies that eventually evolve towards the mid infrared (see Fig.\ \ref{fig3}a), whereas the size range here explored yields tunable visible and mid-infrared excitations.

\section{Conclusions}

The extraordinary second- and third-harmonic generation reported above warrants further exploration of nonlinear optical phenomena in doped nanographene. Other morphologies apart from nanotriangles should yield similar high levels of nonlinear response, particularly when their edges are predominantly armchair. Graphene nanoislands with sizes comparable to those considered here have already been fabricated using various methods \cite{LYG11, SLL12, KHK12}, although they lack precise control over size and shape, which limits their applicability to nonlinear photonic technologies. Alternatively, a bottom-up approach based upon chemical self-asembly of molecular precursors provides better degree of control over the sizes and edge configurations \cite{WPM07, FPM09, BLZ12}. In practical devices, electrical doping can be introduced in the nanoislands through a transparent substrate \cite{paper212}, or by placing them close to a contact, from which electrons can be tunneled. Arrayed nanoislands can transform a substantial fraction of the incident light energy into nonlinear harmonics (see Fig.\ \ref{figS7} in the Appendix). Our results indicate that a relatively small amount of charge transferred to a graphene nanoisland can facilitate a dramatic increase in the magnitude of the nonlinear polarizability. This suggests the use of graphene nanoislands for the development of nanometer-sized optoelectronic switches and modulators, as well as for the detection of minute amounts of analytes through their charge transfer to the graphene.

\section{Methods}

We describe the low-energy ($<3\,$eV) optical response of graphene nanoislands within a density-matrix approach, using a tight-binding model for the $\pi$-band electronic structure. One-electron states $|\varphi_j\rangle$ are obtained by assuming a single $p$ orbital per carbon site, oriented perpendicular to the graphene plane, with a hopping energy of $2.8\,$eV between nearest neighbors \cite{W1947,CGP09}. In the spirit of the mean-field approximation \cite{HL1970}, a single-particle density matrix is constructed as $\rho=\sum_j\tilde{\rho}_{jj'}|\varphi_j\rangle\langle\varphi_{j'}|$, where $\tilde{\rho}_{jj'}$ are time-dependent numbers. An incoherent Fermi-Dirac distribution of occupation numbers $f_j$ is assumed in the unperturbed state, characterized by a density matrix $\tilde{\rho}^0_{jj'}=\delta_{jj'}f_j$, whereas the time evolution under external illumination is governed by the equation of motion
\begin{equation}\label{rho_eom1}
\frac{\partial \rho}{\partial t} = -\frac{i}{\hbar} \left[ H, \rho \right] - \frac{1}{2\tau} \left( \rho - \rho^{0} \right).
\end{equation}
The last term of Eq.\ (\ref{rho_eom1}) describes inelastic losses at a phenomenological decay rate $1/\tau$, where the factor of $1/2$ accounts for the fact that we are relaxing to the ground state instead of the local thermal equilibrium state \cite{M1970}. We set $\hbar\tau^{-1}=10\,$meV throughout this work, corresponding to a conservative Drude-model graphene mobility $\mu\approx1200\,$cm$^2/($V\,s$)$ for a characteristic doping carrier density $4\times10^{13}\,$cm$^{-2}$ (i.e., one charge carrier per every 100 carbon atoms), and note that even higher decay rates still produce large nonlinearities (see Fig.\ \ref{figS3} in the Appendix). The system Hamiltonian $H=H_{\rm TB}-e\phi$ consists of the tight-binding part $H_{\rm TB}$ (i.e., nearest-neighbors hopping) and the interaction with the self-consistent electric potential $\phi$, which is in turn the sum of external and induced potentials. The latter is simply taken as the Hartree potential produced by the perturbed electron density, while the former is related to the incident electric field $\Eb(t)$ as $-\rb\cdot\Eb(t)$. The induced dipole moment is then calculated from the diagonal elements of the density matrix in the carbon-site representation as $\pb(t)=-2e\sum_l\big[\rho_{ll}(t)-\rho^0_{ll}\big]\Rb_l$, where the factor of 2 accounts for spin degeneracy and $\Rb_l$ runs over carbon sites. Under continuous wave illumination ($\Eb(t)=E_0\,\hat{\bf e}\,\ee^{-\ii\omega t}+{\rm c.c.}$), we consider the harmonics contributions to the dipole moment, $\pb(t)=\sum_s \alpha^{(s)}\,(E_0)^s\,\ee^{-\ii s\omega t}+{\rm c.c.}$, which implicitly defines the linear ($s=1$) and nonlinear ($s>1$) polarizabilities $\alpha^{(s)}$. In particular, we focus here on the second- ($s=2$) and third-harmonic ($s=3$) response. We use two different methods to solve Eq.\ (\ref{rho_eom1}) and find $\alpha^{(s)}$: direct time-domain numerical integration and a perturbative approach. We find both of these methods to be in excellent quantitative agreement when the light intensity is sufficiently low. The former also allows us to simulate the response to short light pulses for arbitrarily large intensity. Incidentally, the perturtative approach is a nontrivial extension of the linear random-phase approximation method already reported for graphene \cite{paper183}, and it requires $s$ times as long to compute all harmonics up to order $s$. More details on both of these methods are offered in the Appendix.

\acknowledgments

This work has been supported in part by the European Commission (Graphene Flagship CNECT-ICT-604391 and FP7
-ICT-2013-613024-GRASP).

%%%%%%%%%%%%%%%%%%%%%%%%%%%%%%%%%%%%%%%%%%%%%%%%%%%%%%%%%%%%%
%%%%%%%%%%%%%%%%%%%%%%%%% ------ APPENDIX ------ %%%%%%%%%%%%%%%%%%%%%%%%%
%%%%%%%%%%%%%%%%%%%%%%%%%%%%%%%%%%%%%%%%%%%%%%%%%%%%%%%%%%%%%

\appendix
\appendixpage
\addappheadtotoc

We provide details on the simulation methods used to calculate the optical properties of nanographene. Specifically, we rely on a single-particle density matrix formulation based upon a tight-binding representation of the electronic states. We discuss a time-domain approach to the problem, as well as a perturbative solution method.  We also compile a brief tutorial on nonlinear polarizability units, an analytical anharmonic model of the nonlocal response, additional time-domain numerical simulations, a quantitative picture of the nonlinearities in nanographenes, an analysis of the dependence on the electron relaxation time, an evaluation of the role of defects and edges in the optical response, and simulations of nanoisland arrays.

\section{Theoretical model and methods of solution}

We investigate the linear and nonlinear optical response of doped graphene nanoislands, with emphasis on enhanced nonlinearities arising when the involved optical frequencies are tuned to the graphene plasmons. The latter correspond to light wavelengths that are much larger than the size of the islands \cite{paper176}, and therefore, we characterize their optical response by the induced dipole moment $\textbf{p}$. In our numerical simulations, we consider light incident along the direction normal to the carbon plane, which we take as $\zz$. As we are interested in the enhanced optical response due to plasmons, armchair-edged nanoislands are preferred because they support intense and highly tunable modes, whereas zigzag-edged islands host near-zero-energy electronic states that are detrimental to the strength and tunability of optical excitations \cite{paper214} (see Sec.\ \ref{zigzagsection} below). We concentrate on armchair nanotriangles lying on the $x$-$y$ plane and oriented as illustrated in Fig.\ \ref{fig1}. However, the methods that we introduce below can be directly applied to arbitrarily shaped nanographene.

\subsection{Density-matrix approach to the nonlinear optical response}

We model the optical response using a single-particle density matrix approach, assuming that only $\pi$-band valence electrons are contributing and expanding them in a basis set formed by the 2p carbon orbitals oriented perpendicular to the graphene plane, with one spin-degenerate state $|l\rangle$ per atomic site $\Rb_l=(x_l,y_l)$. We adopt a tight-binding description \cite{W1947,CGP09} in which the unperturbed system is characterized by a Hamiltonian $H_{\rm TB}$ of matrix elements $\langle l|H_{\rm TB}|l'\rangle=-h\delta_{\langle l,l' \rangle}$, where $h=2.8$\,eV is the hopping energy, while $\delta_{\langle l,l' \rangle}$ is 1 if $l$ and $l'$ are nearest-neighbour carbon sites and 0 otherwise. Upon diagonalization of $H_{\rm TB}$, we obtain single-electron states of energies $\hbar\varepsilon_j$ that can be expressed as
\begin{equation}\label{jstate}
|\varphi_j\rangle=\sum_l a_{jl}|l\rangle,
\end{equation}
where the expansion coefficients $a_{jl}$ are real and give the amplitude of orbitals $|l\rangle$ in states $j$. These states are orthonormal ($\sum_l a_{jl}a_{j'l}=\delta_{jj'}$) and form a complete set ($\sum_j a_{jl}a_{jl'}=\delta_{ll'}$). In what follows, we use indices $l$ to label carbon sites and $j$ for single-electron states.

We describe the electronic state of a graphene nanostructure through its single-particle density matrix $\rho$
\begin{equation}\label{rhojjll}
\rho=\sum_{ll'} \rho_{ll'} |l\rangle\langle l'|=\sum_{jj'} \tilde{\rho}_{jj'} |\varphi_j\rangle\langle \varphi_{j'}|,
\nonumber
\end{equation}
where $\rho_{ll'}$ ($\tilde{\rho}_{jj'}$) are time-dependent matrix elements in the site (state) representation. We can move between these two representations using the relations $\tilde{\rho}_{jj'}=\sum_{ll'}a_{jl}a_{j'l'}\rho_{ll'}$ and $\rho_{ll'}=\sum_{jj'}a_{jl}a_{j'l'}\tilde{\rho}_{jj'}$, involving the $a_{jl}$ coefficients defined in Eq.\ (\ref{jstate}). Plasmon dynamics are then studied by solving the equation of motion \cite{HL1970}
\begin{equation}\label{rho_eom}
\frac{\partial \rho}{\partial t} = -\frac{\ii}{\hbar} \left[ H, \rho \right] - \frac{1}{2\tau} \big( \rho - \rho^{0} \big),
\end{equation}
where
\begin{equation}\label{Htot}
H=H_{\rm TB}-e\phi
\end{equation}
is the system Hamiltonian, $\phi$ is the total potential acting on the graphene island, and $\rho^{0}$ is the equilibrium density matrix at time $t=-\infty$ (i.e., before any interaction with external fields), to which the system relaxes at a rate $\tau^{-1}$. We construct $\tilde{\rho}^{0}_{jj'}=f_j\delta_{jj'}$ from the incoherent filling of electron states according to the Fermi-Dirac-distribution occupation numbers $f_j$ \cite{HL1970}. Although this relaxation approximation is unable to conserve local electron density \cite{M1970}, it should provide an accurate description for optical field components oscillating at frequencies $\omega\gg\tau^{-1}$. This is the case of plasmons in high-quality doped graphene, for which we assume a realistic phenomenological inelastic relaxation rate $\hbar\tau^{-1}=10$\,meV throughout this work, unless otherwise stated.

It should be noted that the nonlinearity arises from the induced part of the potential, which produces a quadratic dependence on $\rho$ in the right-hand side of Eq.\ (\ref{rho_eom}). We obtain linear and nonlinear nanographene polarizabilities by numerically solving Eq.\ (\ref{rho_eom}) using either time-domain or perturbative methods, as outlined in the following sections. In particular, results presented in Figs.\ \ref{fig1}, \ref{fig2}(d-f), \ref{figS1}, and \ref{figS5} are obtained from time-domain simulations, while the results of Figs.\ \ref{fig2}(a-c), \ref{fig3}-\ref{fig5}, \ref{figS2}-\ref{figS4}, \ref{figS6}, and \ref{figS7} are calculated using the perturbative approach.

%%%%%%%%%%%%%%%%%%%%%%%%%%%%%%%%%%%%%%%%%%%%%%%%%%%%%%%%%%%%%

\subsection{Time-domain approach}

Direct numerical integration of the equation of motion (\ref{rho_eom}) constitutes an intuitive method of solution, for which it is convenient to express it in the basis set of carbon site orbitals $|l\rangle$ as
\begin{equation}\label{rho_ll'_eom}
\frac{\partial \rho_{ll'}}{\partial t} = 
-\frac{\ii}{\hbar} \sum_{l''} \left( H_{ll''} \rho_{l''l'} - \rho_{ll''} H_{l''l'} \right) - \frac{1}{2\tau} \left( \rho_{ll'} - \rho^{0}_{ll'} \right).
\end{equation}
The elements of the Hamiltonian (see (Eq.\ (\ref{Htot})) are
\begin{equation}\nonumber
H_{ll'}=-h\delta_{\langle ll' \rangle}-e\delta_{ll'}\phi_l.
\end{equation}
Here, the total potential $\phi$ is diagonal in the site representation and results from the sum of the external potential \[\phi^{\text{ext}}_l=-\textbf{R}_l\cdot\textbf{E}(t),\] where $\Eb(t)$ is the incident electric field, assumed to be uniform along the island, and the self-consistent induced potential, which we model in the Hartree approximation as \cite{HL1970}
\begin{equation}\label{phiind}
\phi^{\text{ind}}_l=-2e\sum_{l'}v_{ll'}\left( \rho_{l'l'} - \rho^{0}_{l'l'} \right)
\end{equation}
after correcting for homogeneous doping in the graphene nanoisland \cite{PGW12}. Here, $-2e(\rho_{l'l'}-\rho^{0}_{l'l'})$ is the induced charge at site $l'$, $v_{ll'}$ gives the spatial dependence of the Coulomb interaction between electrons in orbitals $|l\rangle$ and $|l'\rangle$ \cite{paper183}, and the factor of 2 accounts for electron spin.

The time-dependent elements $\rho_{ll'}$ are calculated by numerically integrating Eq.\ (\ref{rho_ll'_eom}) to yield the induced dipole moment
\begin{equation}\nonumber
\textbf{p}(t)=-2e\sum_l\left(\rho_{ll}-\rho^{0}_{ll}\right)\textbf{R}_l.
\end{equation}
We use this approach to study the response to high-fluence Gaussian light pulses (see Sec.\ \ref{Intensitydependence} below and Fig.\ \ref{fig2}(d-f)).

Additionally, we are interested in computing second- and third-harmonic generation (SHG and THG) upon continuous-wave (CW) illumination. Accordingly, we write the incident field as
\begin{equation}\label{EE0}
\Eb(t)=E_0\,\ee^{-\ii\omega t}\,\hat{\textbf{e}}+{\rm c.c.},
\end{equation}
where $\hat{\textbf{e}}$ is the polarization vector, which we take for simplicity along a high-symmetry direction of the system, so that the induced dipole is also along $\hat{\textbf{e}}$ (this is the case for polarization either parallel or perpendicular to one of the sides of an equilateral triangular island). For moderate light intensities, the leading contribution to the dipole oscillating at the $s^{\rm th}$ harmonic is defined as
\begin{equation}\nonumber
p^{(s)}(t)=\alpha^{(s)}\,E_0^s\,\ee^{-\ii s\omega t}+{\rm c.c.}, \nonumber
\end{equation}
where $\alpha^{(s)}$ is the $s$ order polarizability. We calculate the latter by separating it from all other constituent terms in the induced dipole $p(t)$ upon Fourier transformation of a single optical cycle. More precisely, from the above definition of $p^{(s)}(t)$ it follows that
\begin{equation}\nonumber
\alpha^{(s)}(\omega)=\frac{\omega}{2\pi (E_0)^s}\int^{t_1}_{t_1-2\pi/\omega}p(t)\,\ee^{\ii s\omega t}dt, \nonumber
\end{equation}
where we assume the system to have evolved to a steady state at a time $t_1\gg\tau$ after starting the simulation. We use this procedure to compute the frequency-dependent SHG and THG polarizabilities represented in some of the above figures and in Fig.\ \ref{figS5} (see below).

%%%%%%%%%%%%%%%%%%%%%%%%%%%%%%%%%%%%%%%%%%%%%%%%%%%%%%%%%%%%%

\subsection{Perturbative approach}

An iterative solution of Eq.\ (\ref{rho_eom}) is possible under CW illumination. This is facilitated by writing it in the state representation as
\begin{align}
\frac{\partial \tilde{\rho}_{jj'}}{\partial t} = &-\ii\left(\varepsilon_j - \varepsilon_{j'}\right)\tilde{\rho}_{jj'}
\label{dmjj}\\ &+
\frac{\ii e}{\hbar}\sum_{l,l'} \left( \phi_l - \phi_{l'} \right)a_{jl}a_{j'l'}\rho_{ll'} -\frac{1}{2\tau}\left( \tilde{\rho}_{jj'}-\tilde{\rho}^{0}_{jj'}\right),
\nonumber
\end{align}
where we have used $H_{\rm TB}|j\rangle=\hbar\varepsilon_j|j\rangle$, and the interaction potential term has been transformed using the coefficients $a_{jl}$ of Eq.\ (\ref{jstate}). We then expand the density matrix as
\begin{equation}\label{rhons}
\rho=\sum_{n,s} \rho^{ns} \ee^{-\ii s\omega t},
\end{equation}
where $n=1,2,3,...$ labels the perturbation order (i.e., terms proportional to $(E_0)^n$, see Eq.\ (\ref{EE0})), while $s$ indicates the harmonic. We use the property $\tilde{\rho}_{jj'}^{ns}=\left(\tilde{\rho}_{j'j}^{n,-s}\right)^*$ to reduce the computation time and storage demand roughly by a factor of 2. To $0^{\rm th}$ order in $E_0$, Eq.\ (\ref{dmjj}) is trivially satisfied with $\rho^{ns}=\delta_{s,0}\rho^0$. The external potential contributes to first order ($n=1$) with $s=\pm1$ components only. We obtain the solution at higher orders by inserting Eq.\ (\ref{rhons}) into Eq.\ (\ref{dmjj}) and identifying terms with the same $\ee^{-\ii s\omega t}$ dependence on both sides of the equation. Clearly, we have $|s|\leq n$ by construction. At order $n\ge1$, we find
\begin{equation}\label{rhons2}
\tilde{\rho}^{ns}_{jj'}=-\frac{e}{\hbar}\sum_{l,l'}\frac{\left(\phi^{ns}_l-\phi^{ns}_{l'}\right)a_{jl}a_{j'l'}}
{s\omega+\ii/2\tau-\left(\varepsilon_j-\varepsilon_{j'}\right)}\rho^{0}_{ll'}+\eta^{ns}_{jj'},
\end{equation}
where
\begin{equation}\label{etajj}
\eta^{ns}_{jj'}=-\frac{e}{\hbar}\sum^{n-1}_{n'=1}\sum^{n'}_{s'=-n'}\sum_{l,l'}
\frac{\left(\phi^{n's'}_l-\phi^{n's'}_{l'}\right)a_{jl}a_{j'l'}}
{s\omega+\ii/2\tau-\left(\varepsilon_j-\varepsilon_{j'}\right)}\rho^{n-n',s-s'}_{ll'},
\end{equation}
while
\begin{equation}\label{phinsrho}
\phi^{ns}_{l}=\phi^{\text{ext}}_l\delta_{n,1}(\delta_{s,-1}+\delta_{s,1})-2e\sum_{l'} v_{ll'} \rho^{ns}_{l'l'}
\end{equation}
is the contribution to the harmonic $s$ of the total potential at order $n$. In Eq.\ (\ref{rhons2}), the first term on the right-hand side has a linear dependence on $\rho^{ns}$ through the induced part of $\phi^{ns}$ (i.e., the sum in Eq.\ (\ref{phinsrho})), whereas we have separated the dependence on lower perturbation orders in $\eta^{ns}_{jj'}$. At each order $n$ we are thus dealing with a self-consistent system in $\phi^{ns}$, which we handle in a way similar to the random-phase approximation (RPA) formalism in linear response theory \cite{HL1970}. We proceed by first using the identity $\tilde{\rho}_{ll'}^0=\sum_{jj'}a_{jl}a_{j'l'}\tilde{\rho}_{jj'}^0=\sum_ja_{jl}a_{jl'}f_j$ in the sum of Eq.\ (\ref{rhons2}), and then moving from state to site representation to obtain the diagonal density-matrix elements as
\begin{equation}\label{rhonsll2}
\rho^{ns}_{ll}=\frac{-1}{2e}\sum_{l'}\chi^{0,(s)}_{ll'}\phi^{ns}_{l'}+\sum_{jj'}a_{jl}a_{j'l}\eta^{ns}_{jj'},
\end{equation}
where
\begin{equation}\label{chi0}
\chi^{0,(s)}_{ll'}=\frac{2e^2}{\hbar}\sum_{jj'}\left(f_{j'}-f_j\right)\frac{a_{jl}a_{j'l}a_{jl'}a_{j'l'}}{s\omega+\ii/2\tau-\left(\varepsilon_j-\varepsilon_{j'}\right)}
\end{equation} 
is the noninteracting RPA susceptibility at frequency $s\omega$.

In summary, each new iteration order $n$ is computed from the results of previous orders as follows:
\begin{enumerate}
\item We first calculate $\eta^{ns}_{jj'}$ using Eq.\ (\ref{etajj}).
\item We then combine Eqs.\ (\ref{phinsrho}) and (\ref{rhonsll2}) to find a self-consistent equation for $\phi_l^{ns}$, which reduces in matrix form to
\begin{equation}\nonumber
\phi^{ns}=(1-v\cdot\chi^{0,(s)})^{-1}\cdot\beta^{ns},
\end{equation}
using site labels $l$ as matrix indices and having defined
\begin{equation}\nonumber
\beta^{ns}_l=\phi^{\text{ext}}_l\delta_{n,1}(\delta_{s,-1}+\delta_{s,1})-2e\sum_{l'jj'}v_{ll'}a_{jl}a_{j'l}\eta_{jj'}^{ns}.
\end{equation}
\item We use the calculated values of $\eta_{jj'}^{ns}$ and $\phi^{ns}_l$ to obtain $\rho^{ns}_{ll}$ using Eq.\ (\ref{rhonsll2}), and from here the induced charge at site $l$ at order $n$ associated with harmonic $s$ as $\rho^{\rm ind}_l=-2e\rho^{ns}_{ll}$. Incidentally, we can also calculate the full density matrix $\tilde{\rho}_{jj'}$ using Eq.\ (\ref{rhons2}).
\item Finally, the polarizability of order $s$ is calculated from
\begin{equation}
\alpha^{(s)}=-\frac{2e}{E_0^s}\sum_l\rho^{ss}_{ll}\;\Rb_l \cdot \hat{\textbf{e}}
\end{equation}
upon iteration of this procedure up to order $n=s$.
\end{enumerate}

%%%%%%%%%%%%%%%%%%%%%%%%%%%%%%%%%%%%%%%%%%%%%%%%%%%%%%%%%%%%%

\begin{table*}
\begin{center}
\begin{tabular}{l|ccc}
\hline \\     Units       &\;\; esu  \;\;&\;\; SI \;\;&\;\; a.u.
\\ \\ \hline \\
$\alpha^{(1)}$\;\;\;\;\;\;\;
&\;\; $1\,\text{cm}^3$                                    \;\;
&\;\; $1.113\times10^{-16}\,\text{C}^2\,\text{m}^2\,\text{J}^{-1}$ \;\;
&\;\; $6.748\times10^{24}$
\\ \\
$\alpha^{(2)}$\;\;\;\;\;\;\;
&\;\; $1\,\text{cm}^5/\text{statC}$                                    \;\;
&\;\; $3.711\times10^{-21}\,\text{C}^3\,\text{m}^3\,\text{J}^{-2}$ \;\;
&\;\; $1.157\times10^{32}$
\\ \\
$\alpha^{(3)}$\;\;\;\;\;\;\;
&\;\; $1\,\text{cm}^7/\text{statC}^2$                                    \;\;
&\;\; $1.238\times10^{-25}\,\text{C}^4\,\text{m}^4\,\text{J}^{-3}$ \;\;
&\;\; $1.985\times10^{39}$
\\ \\ \hline
\end{tabular}
\end{center}
\caption{Conversion factors between electrostatic units (esu), international system units (SI), and atomic units (a.u.) for linear, SHG, and THG polarizabilities. The equivalent of 1\,esu is expressed in SI and a.u. units in each case.}
\label{table1}
\end{table*}

\section{Nonlinear polarizability units}

In the literature, the nonlinear polarizabilities $\alpha^{(2)}$ and $\alpha^{(3)}$ are commonly reported in Gaussian electrostatic units (esu) \cite{B08_3,KRM94}, with length in cm and charge in statcoulomb ($1\,$statC$\,=4\pi c_0\,$C, where $c_0=2,997,924,580$ is the speed of light expressed in m$/$s), whereas many theoretical studies use atomic units (a.u., with $e=\hbar=m_e=1$). The conversion factors between esu, SI, and a.u. are given in Table\ \ref{table1}.

Throughout this work we adopt the common convention of referring to the esu units of $\alpha^{(1)}$ ($\text{cm}^3$), $\alpha^{(2)}$ ($\text{cm}^5$/statC), and $\alpha^{(3)}$ ($\text{cm}^7$/$\text{statC}^2$) simply as ``esu'' \cite{KRM94}. For completeness, in Fig.\ \ref{fig5} we compare our simulated nonlinear polarizabilities $\alpha^{(2)}$ and $\alpha^{(3)}$ with previously reported measurements for noble metal nanoparticles, and offer both esu and SI values, as obtained by using the noted table. In particular, we show experimental values of $|\alpha^{(2)}|$ acquired through hyper-Rayleigh scattering of various types of gold and silver nanoparticles in aqueous suspensions \cite{VLH98, RBB07, GBG99, RDB10, SSN09}. Additionally, we show values of the third-order susceptibility $|\chi^{(3)}|$ obtained from degenerate four-wave mixing measurements of glasses doped with silver and copper nanoparticles \cite{UKO94} and from THG measurements of dispersed silver colloids on quartz \cite{LTY06}, converted into $|\alpha^{(3)}|$ upon multiplication by the particle volume.

%%%%%%%%%%%%%%%%%%%%%%%%%%%%%%%%%%%%%%%%%%%%%%%%%%%%%%%%%%%%%

\section{Anharmonic oscillator model}

The intensity dependence of the first-harmonic polarizability can be qualitatively described by a classical model for damped and optically driven particles of mass $m_e$ subjected to an effective anharmonic potential \cite{FT1980, RB1985}. The equation of motion for such an oscillator can be written as
\begin{equation}\label{aho_eom}
m_e\ddot{x}+m_e\tau^{-1}\dot{x}=-f\,e\,E(t)+\partial_xU(x),
\end{equation}
where $f$ quantifies the coupling strength to the driving electric field $E(t)=E_0e^{-\ii\omega t}+\text{c.c.}$, while $U(x)=-(m_e\omega_0^2/2) x^2-(m_ea/4)x^4$ is the anharmonic potential, characterized by a resonance frequency $\omega_0$ and a nonlinear coefficient $a$. The solution for the particle position $x(t)$ can be expressed in harmonics of the driving field as
\begin{equation}\nonumber
x(t)=\sum_{s=1}^{\infty} x^{(s)}(\omega)e^{-\ii s\omega t}+{\rm c.c.}
\end{equation}
If we neglect terms of order $s>1$ in the above expansion, Eq.\ (\ref{aho_eom}) leads to
\begin{equation}\label{aho_x1}
-3a|x^{(1)}|^2x^{(1)}+\left[\omega\left(\omega+i\tau^{-1}\right)-\omega_0^2\right]x^{(1)}=feE_0/m_e
\end{equation}
for the first-harmonic amplitude $x^{(1)}$.
The cubic equation (\ref{aho_x1}) can be used to explain the results obtained from our time-domain simulations for the linear polarizability of nanographene under intense CW illumination. In Fig.\ \ref{fig2}d, we successfully fit $\alpha^{(1)}=-ex^{(1)}/E_0$ to this model for one of the islands by taking $\hbar\omega_0=0.68\,$eV, $f=1.77$, and $a=-(4.7+0.3\,\ii)\times10^{47}\,$m$^{-2}$s$^{-2}$.
It should be noted that the anharmonic oscillator model described here must be considered as a phenomenological tool, used simply to qualitatively illustrate the magnitude of the nonlinear shift observed in the linear spectra for illumination by intense, CW fields.

%%%%%%%%%%%%%%%%%%%%%%%%%%%%%%%%%%%%%%%%%%%%%%%%%%%%%%%%%%%%%

\begin{figure*}
\includegraphics[width=1\textwidth]{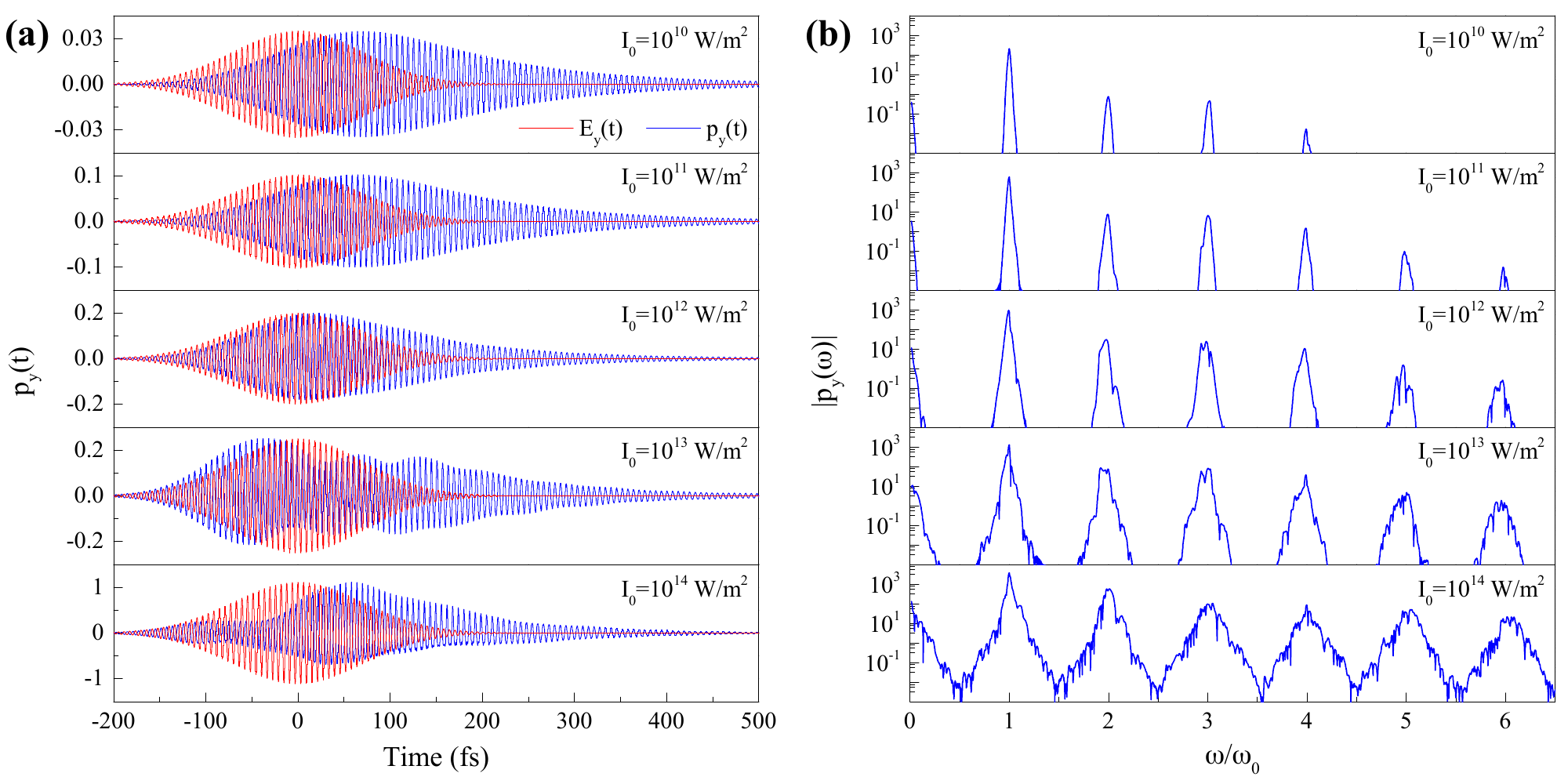}
\caption{\label{figS1} Time-domain simulations of an armchair graphene nanotriangle excited by a 166\,fs FWHM Gaussian pulse with various peak intensities $I_0$. We show the temporal evolution (left, blue curves) and the spectral dependence (right) of the induced dipole moment. The island consists of 330 carbon atoms and is doped with two electrons. The central frequency of the pulse is tuned to the $0.68$\,eV plasmon (see Fig.\ \ref{fig2}a). The polarization direction is perpendicular to one of the triangle sides.}
\label{figS1}
\end{figure*}

\section{Intensity-dependence in the time-domain}
\label{Intensitydependence}

We present in Fig.\ \ref{figS1}a numerical simulations for the temporal evolution of the induced dipole moment in a doped graphene nanoisland under excitation by 166\,fs FWHM laser Gaussian pulses of high intensity, as obtained by direct numerical integration of Eq.\ (\ref{rho_ll'_eom}). For relatively small pulse fluence, the polarization displays a temporally delayed Gaussian-like profile. At higher pulse fluences, the anharmonicitiy of nanographene leads to a strongly asymmetric response. The frequency spectrum of the induced polarization (Fig.\ \ref{figS1}b) reveals the enhanced excitation of high harmonics with intense pulses.

%%%%%%%%%%%%%%%%%%%%%%%%%%%%%%%%%%%%%%%%%%%%%%%%%%%%%%%%%%%%%

\begin{figure*}
\includegraphics[width=0.95\textwidth]{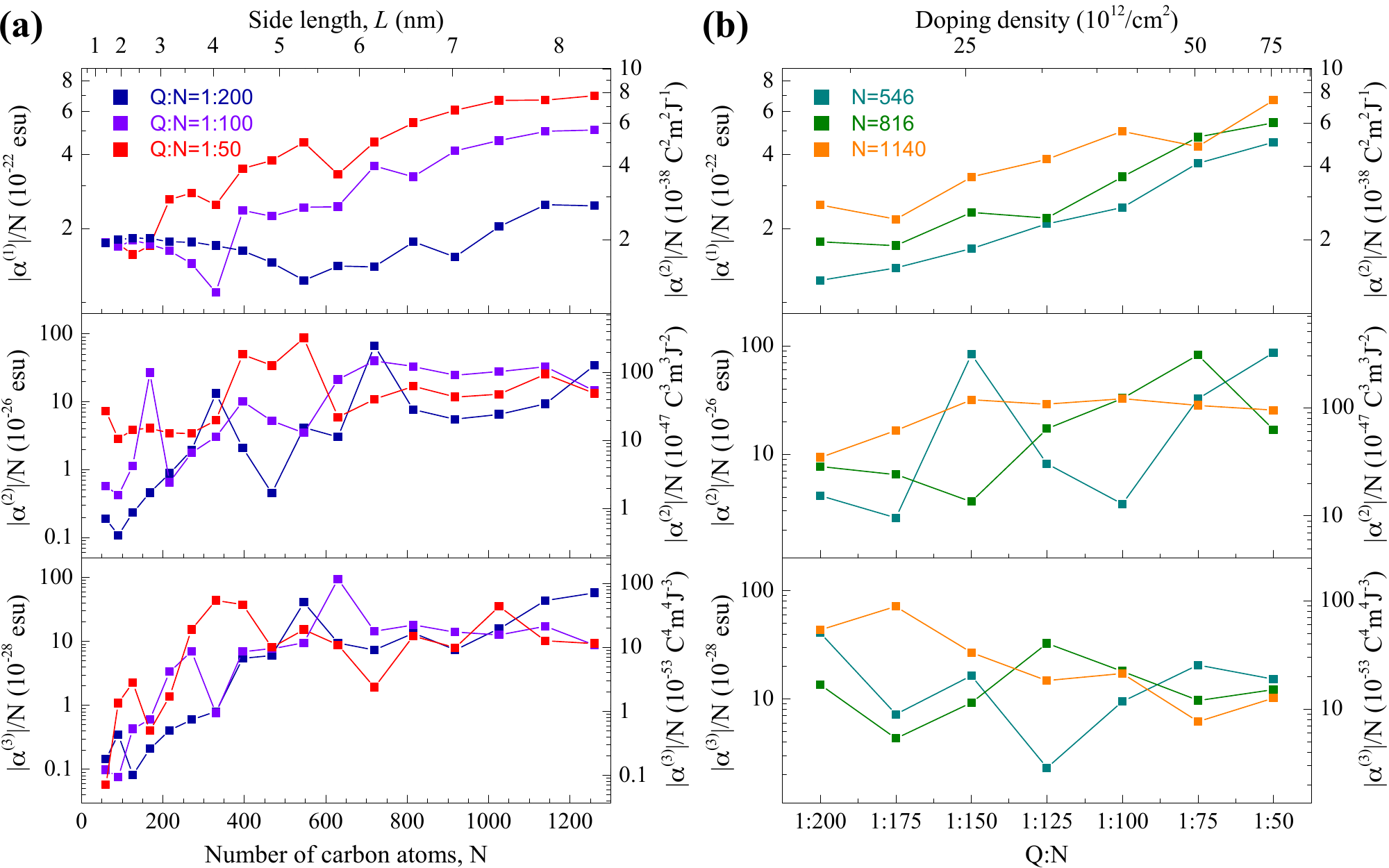}
\caption{Linear and nonlinear peak polarizabilities per carbon atom at the dominant plasmon features of graphene armchair nanotriangles as a function of their size (a) and doping level (b) for selected doping levels and sizes, respectively. These data correspond to the same conditions as in Fig.\ \ref{fig3}.}
\label{figS2}
\end{figure*}

\section{Quantitative analysis of optical nonlinearities}

In Fig.\ \ref{figS2} we show the specific values for the linear and nonlinear peak polarizabilities used to produce Fig.\ \ref{fig3}, normalized to the number of carbon atoms in each nanotriangle. The polarizability per carbon atom is a sensible metric for identifying anomolously large nonlinearities in specific nanotriangles, as it effectively removes the size dependence of the induced dipole. An estimate for the nonlinear susceptibilities $\chi^{(2)}$ and $\chi^{(3)}$ for doped nanographene can be obtained from the polarizability per atom (denoted here as $\tilde{\alpha}^{(s)}$ for the harmonic $s$) by using the relation $\chi^{(s)}=\tilde{\alpha}^{(s)}n_C/d_{gr}$, where $n_C=3.8\times10^{15}$\,cm$^{-2}$ is the density of carbon atoms in bulk graphene and $d_{gr}\simeq3.3\times10^{-8}$\,cm is the effective thickness of a graphene layer. Using this approach, the results presented in Fig.\ \ref{figS2} yield peak nonlinear susceptibilities $|\chi^{(3)}|$ in the $10^{-7}$--$10^{-3}$\,esu range, to be compared with those measured for bulk graphene \cite{HHM10} ($|\chi^{(3)}|\sim10^{-7}$\,esu, or equivalently, $\tilde{\alpha}^{(3)}\sim10^{-30}$\,esu).

%%%%%%%%%%%%%%%%%%%%%%%%%%%%%%%%%%%%%%%%%%%%%%%%%%%%%%%%%%%%%

\begin{figure*}
\includegraphics[width=0.7\textwidth]{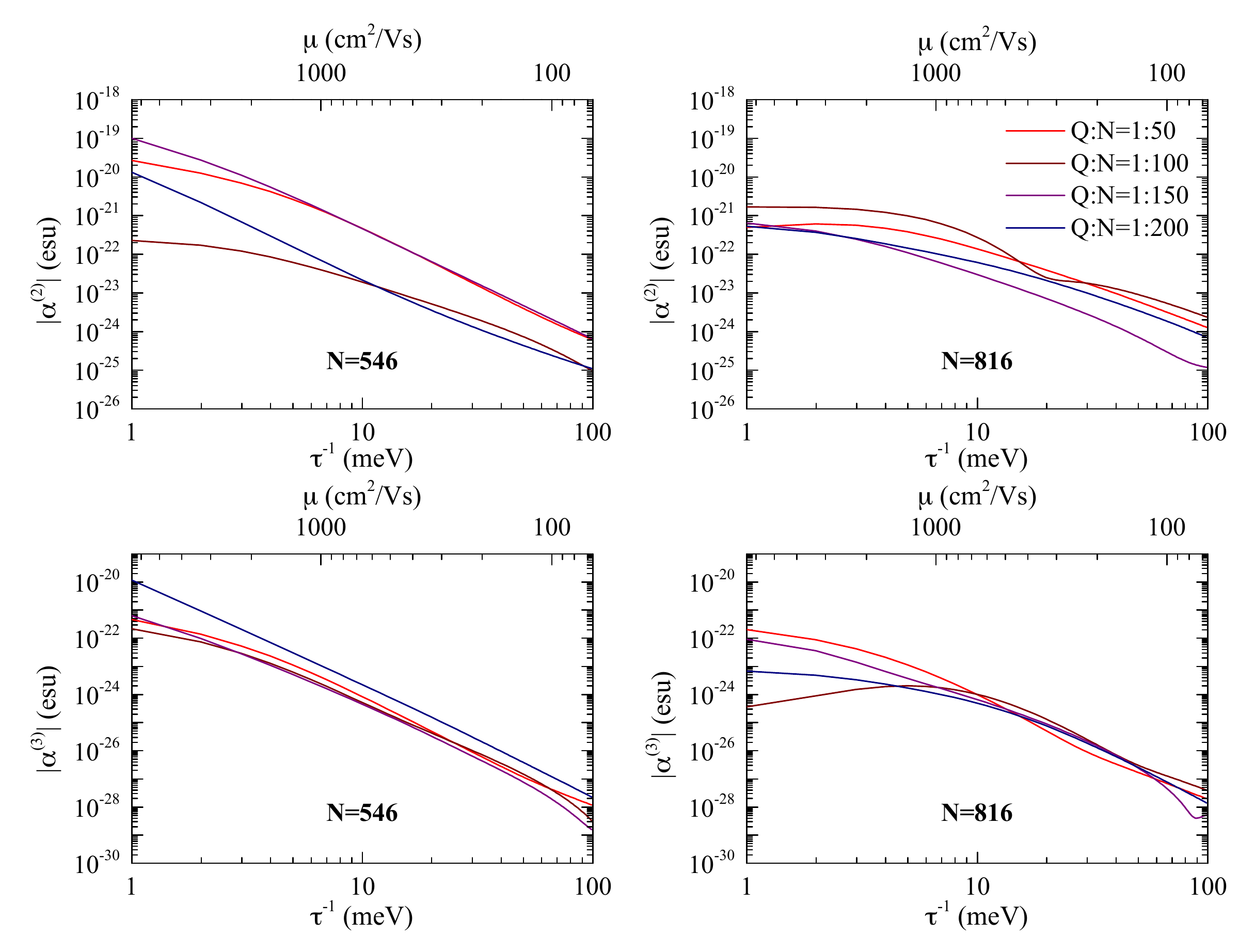}
\caption{Relaxation rate dependence of the maximum SHG and THG polarizabilities for two different graphene nanoislands and various doping levels (see legend in upper right panel). For comparison, we show in the upper horizontal axes the carrier mobilities corresponding to the values of the relaxation time $\tau$ in the lower horizontal axis, calculated in the Drude model as $\mu=ev_F^2/\tau E_F$, where $v_F\approx c/300$ is the Fermi velocity in graphene and we use a representative Fermi energy $E_F=1$\,eV.}
\label{figS3} 
\end{figure*}

\section{Effect of the relaxation rate}

The large linear and nonlinear polarizabilities found in the graphene nanoislands depend strongly on the choice of the plasmon relaxation rate $\tau^{-1}$. Throughout this work, except in this section, we take $\hbar\tau^{-1}=10$\,meV, which is comparable to the values estimated from DC impurity-limited mobilities in high-quality graphene \cite{NGM04, NGM05, JBS09} and similar to values reported for graphene used in actual plasmonic studies \cite{paper212}. This is a conservative choice, because chemical synthesis of finite-sized nanoislands should enable fabrication of nanographene with fewer defects and less disorder. Arguably, graphene phonons may also contribute to losses via phonon-plasmon coupling decay channels, in particular for optical phonons near $\sim0.2\,$eV \cite{JBS09}. These modes are essentially connected with the stretching oscillations of C-C bonds, so they are very localized and give rise to relatively dispersionless bands. Therefore, their contribution to inelastic attenuation acts locally and we expect them to act in a similar way as in extended graphene even in small nanoislands.

In order to demonstrate that the extraordinary nonlinear response of the doped graphene nanoislands presented here is robust against larger plasmon decay rates, we show in Fig.\ \ref{figS3} the maximum nonlinear polarizabilities for two nanoislands as the decay rate $\tau^{-1}$ is varied in the $1$--$100\,$meV range, corresponding to decay times $\tau\sim7$--$660\,$fs. We find that even with very high relaxation rates the peak polarizabilities in graphene are competitive with those measured in noble metal nanoparticles, and maintain their advantage of electrostatic tunability.

%%%%%%%%%%%%%%%%%%%%%%%%%%%%%%%%%%%%%%%%%%%%%%%%%%%%%%%%%%%%

\begin{figure*}
\includegraphics[width=0.7\textwidth]{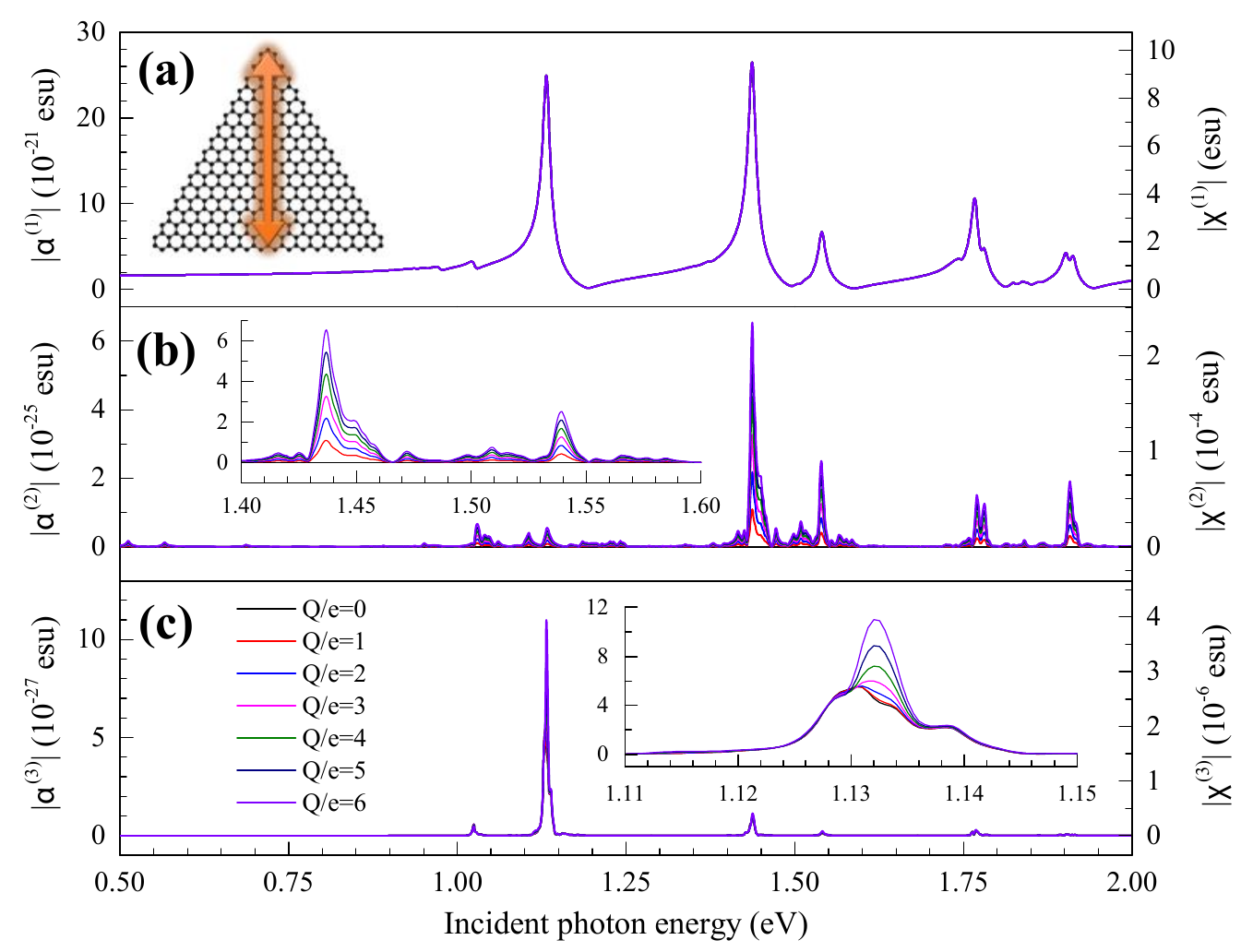}
\caption{We study the linear (a), second-harmonic (b), and third-harmonic (c) polarizabilities of a nanotriangle with zigzag edges for low-intensity continuous illumination, where the electric field polarization is perpendicular to one of the graphene sides. The nanoisland contains $N=321$ carbon atoms, and the number of additional charge carriers is varied between 0 and 6. The insets in (b) and (c) show parts of the second- and third-harmonic polarizability spectra, respectively, in greater detail.}
\label{figS4} 
\end{figure*}

\begin{figure*}
\includegraphics[width=0.7\textwidth]{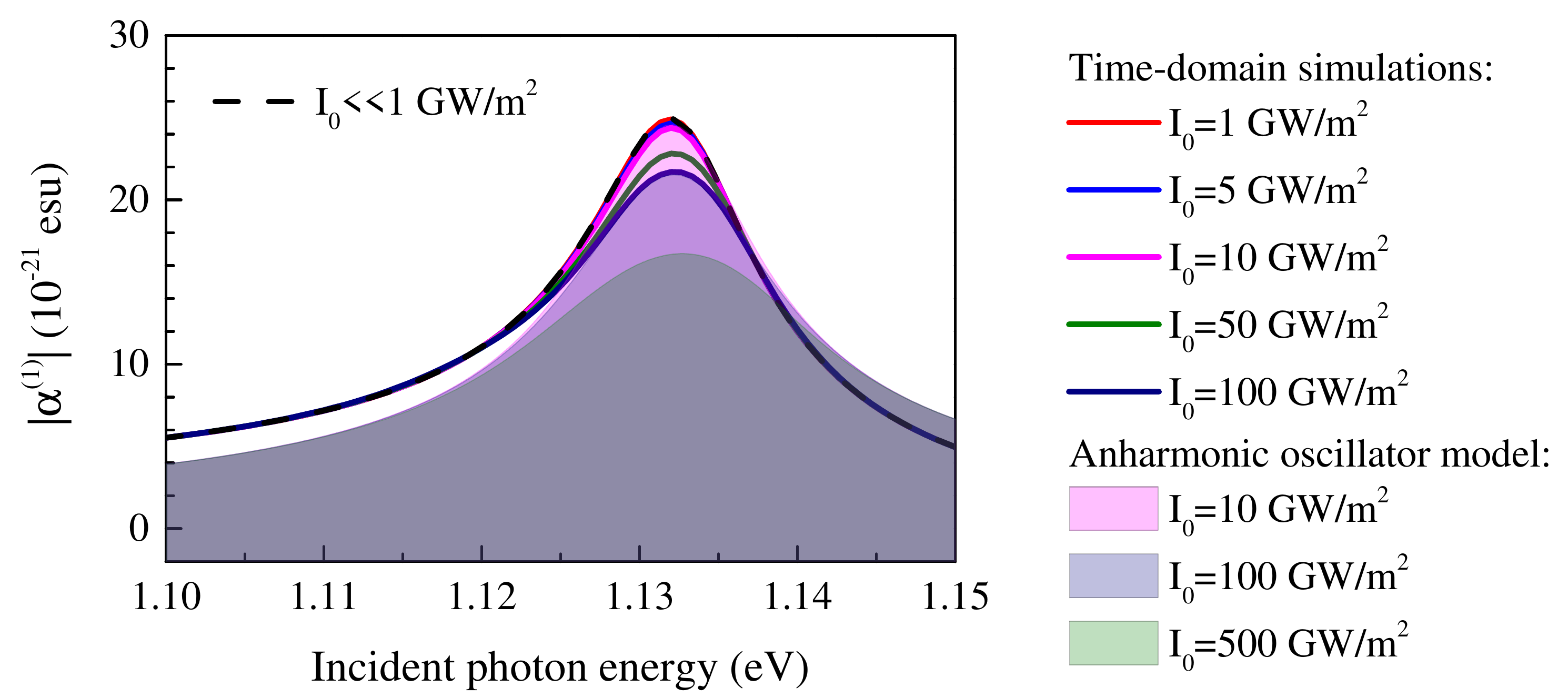}
\caption{Linear polarizability near the lowest-energy dipole plasmon of the nanoisland considered in Fig.\ \ref{figS4}, with doping $Q=2$, calculated for various strong illumination intensities $I_0$ (see legends). The filled curves are obtained using the classical anharmonic oscillator model.}
\label{figS5} 
\end{figure*}

\section{Effect of zigzag edges}
\label{zigzagsection}

The optical response of a graphene nanoisland strongly depends its edge configuration. In particular, previous studies have revealed that zigzag edges are highly detrimental to the tunability of plasmons \cite{paper214}, particularly if the plasmon energy exceeds the Fermi energy, due to the involvement of zero-energy electronic edge states \cite{paper235}. In order to explore the role of zigzag edges on the nonlinear optical response, we show in Fig.\ \ref{figS4} the linear and nonlinear polarizabilities for a zigzag-edged nanotriangle with a similar number of atoms and the same doping conditions as the nanoisland considered in Fig.\ \ref{fig2}. For the range of doping densities considered here, we find that the linear response of the nanoisland remains unchanged, with no traces of tunable plasmons as the doping is increased. Additionally, features in the nonlinear polarizability spectra remain at fixed frequencies, but their amplitudes increase with doping. We note that the maximum SHG and THG polarizabilities are smaller than those of the armchair nanoisland of similar size studied in Fig.\ \ref{fig2}, yet comparable to the nonlinear polarizabilities of noble metal nanoparticles (see Fig.\ \ref{fig5}). To complete the analogy with that figure, we also investigate the intensity dependence of the linear polarizability for the zigzag nanotriangle using time-domain simulations under CW illumination. In Fig.\ \ref{figS5}, we show the lowest-energy dipole plasmon for $Q=2$ doping as the intensity of the incident light is increased. These time-domain simulations are again well mimicked by the classical anharmonic oscillator model given in Eq.\ (\ref{aho_x1}), in this case taking the resonance frequency $\hbar\omega_0=1.132\,$eV, the coupling parameter $f=2.59$, and the nonlinear coefficient $a=(2.6-8.6\ii)\times10^{46}\,$(ms)$^{-2}$.

%%%%%%%%%%%%%%%%%%%%%%%%%%%%%%%%%%%%%%%%%%%%%%%%%%%%%%%%%%%%

\section{Defects in graphene nanoislands}

\begin{figure*}
\includegraphics[width=1\textwidth]{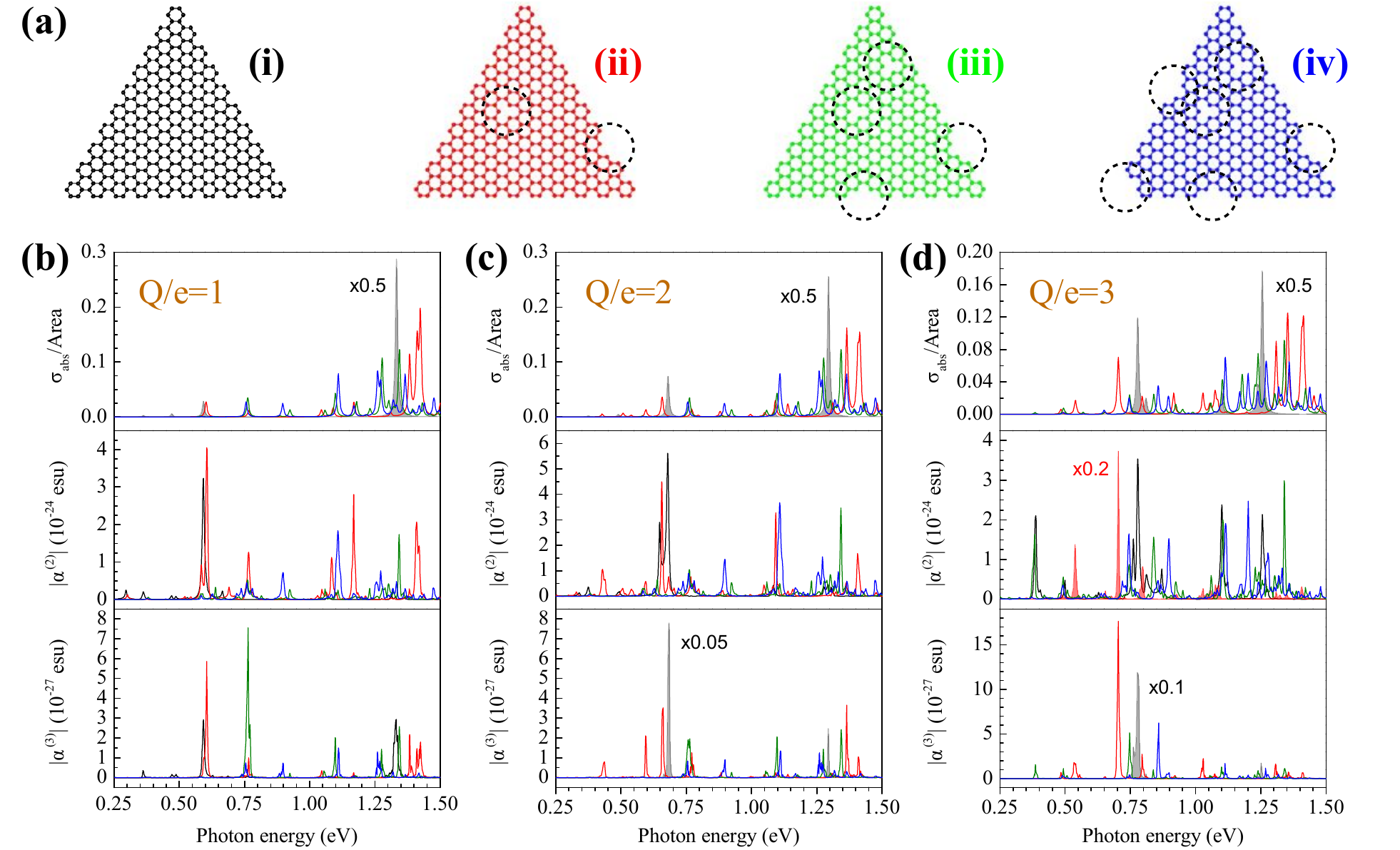}
\caption{Optical response of the $N=330$ triangle considered in Fig.\ \ref{fig1} and \ref{fig2}, studied as defects are added to the structure by randomly removing carbon atoms. (a) Atomic structures of the nanoislands under consideration. (b-d) Spectral dependence of the linear, SHG, and THG polarizabilities for various doping levels, with curve colors chosen to match the atomic structure plots of (a).}
\label{figS6} 
\end{figure*}

The sensitivity of the optical reponse in graphene nanoislands to defects is examined in Fig.\ \ref{figS6}, where the linear and nonlinear polarizabilities of the {\it perfect} armchair nanoisland featured in Figs.\ \ref{fig1} and \ref{fig2} are compared to those of three variations formed by randomly removing carbon atoms from the nanostructure. These defects create additional modes in the linear absorption spectrum, with diminished strength compared to those of the ideal nanoisland. Interestingly, the magnitudes of the second-order polarizability are not significantly decreased in the defective nanoislands. Actually, some of the new modes can produce double-resonance enhancement (see Fig.\ \ref{fig4}), as for example in the structure labeled (ii) in Fig.\ \ref{figS6} with doping $Q=3$.

%%%%%%%%%%%%%%%%%%%%%%%%%%%%%%%%%%%%%%%%%%%%%%%%%%%%%%%%%%%%

\begin{figure*}
\includegraphics[width=1\textwidth]{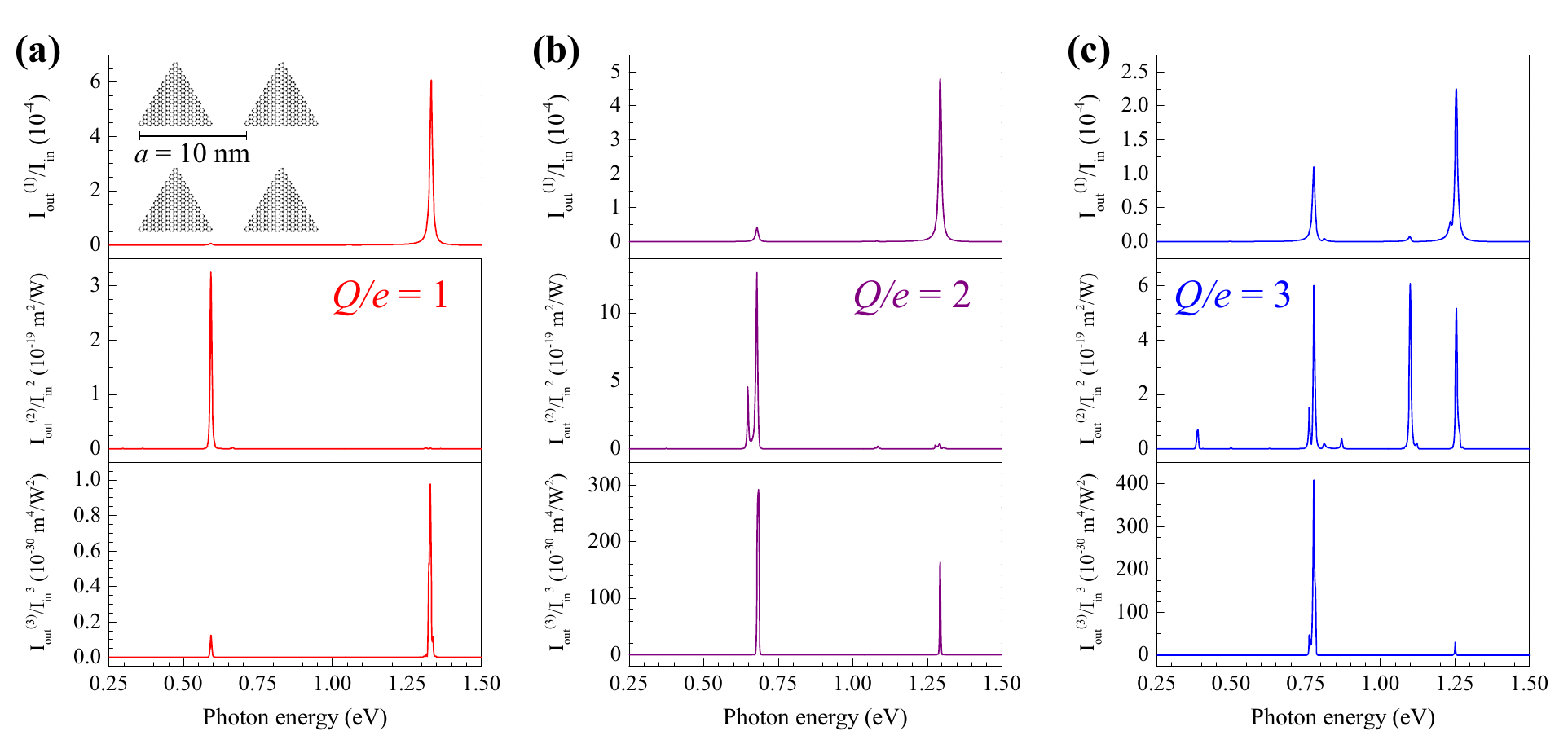}
\caption{Far-field intensity generated by 2D square arrays (period $a=10\,$nm) of graphene nanoislands at the fundamental and higher harmonics of an incident plane-wave field with intensity $I_0$. The doping charge in the nanoislands is $Q=1$, 2, and 3 in  (a), (b), and (c), respectively.}
\label{figS7} 
\end{figure*}

\section{Nanoisland arrays}

The experimental realization of SHG and THG from graphene nanoislands could benefit from using samples containing large numbers of nanoislands. We estimate here the nonlinear signal intensity resulting from an array of nanotriangles illuminated by a light plane wave. For simplicity, we consider a square array under normal incidence conditions. From the response of a dipole array, the linear reflected electric field is given by the expression \cite{paper182}
\begin{equation}
E^{\text{ref}}=\frac{\ii S(\omega)}{1/\alpha^{(1)}(\omega)-G(\omega)}\;E_0,
\end{equation}
where $G(\omega)\approx4.52/a^3+\ii\,[S(\omega)-2\omega^3/3c^3]$ is a lattice sum that accounts for the dipole-dipole interactions between nanoislands, $a$ is the period of the array, and $S(\omega)=2\pi\omega/ca^2$, assuming $\omega a/c\ll1$. Now, we extend this approach to describe higher harmonics in terms of the nonlinear polarizabilities. A straightforward generalization of the methods of Ref.\ \cite{paper182} permits writing the intensity of the $s$ harmonic normalized to the incident intensity $I_0$ as
%\begin{equation}
%\frac{I^{(s)}}{I_0^s}=\left(\frac{2\pi s \omega}{ca^2}\right)^2\left(\frac{8\pi}{c}\right)^{n-1}\left|\frac{\alpha^{(n)}\left[1+\alpha^{(1)}(\omega)G(\omega)\right]^s}{1-\alpha^{(1)}(s\omega)G(s\omega)}\right|^2.
%\end{equation}
\begin{align}
\frac{I^{(s)}}{(I_0)^s}=&\left(\frac{s \omega}{a^2}\right)^2\;\left(\frac{2\pi}{c}\right)^{s+
1}\nonumber\\ 
&\times \left|\frac{\alpha^{(s)}(\omega)}{\left[1-\alpha^{(1)}(s\omega)G(s\omega)\right]\,\left[1-\alpha^{(1)}(\omega)G(\omega)\right]^s}\right|^2.
\nonumber
\end{align}
We show in Fig.\ \ref{figS7} the power emitted towards one side of an array at frequencies $s\omega$ for a period  $a=10\,$nm and islands like those of Fig.\ \ref{fig2}, using the nonlinear polarizabilities reported in that figure. For example, a peak intensity of $1\,$GW/m$^2$ ($100\,$GW/m$^2$) will produce $1.3\,$W/m$^2$ ($13\,$kW/m$^2$) and $0.29\,$mW/m$^2$ ($0.29\,$MW/m$^2$) via SHG and THG, respectively, when the nanoislands are doped with $Q=2$ (Fig.\ \ref{figS7}b).

%\bibliographystyle{achemso}
%\bibliography{../../bibtex/refs}
%\bibliography{../../refs}

\begin{mcitethebibliography}{54}
\providecommand*{\natexlab}[1]{#1}
\providecommand*{\mciteSetBstSublistMode}[1]{}
\providecommand*{\mciteSetBstMaxWidthForm}[2]{}
\providecommand*{\mciteBstWouldAddEndPuncttrue}
  {\def\EndOfBibitem{\unskip.}}
\providecommand*{\mciteBstWouldAddEndPunctfalse}
  {\let\EndOfBibitem\relax}
\providecommand*{\mciteSetBstMidEndSepPunct}[3]{}
\providecommand*{\mciteSetBstSublistLabelBeginEnd}[3]{}
\providecommand*{\EndOfBibitem}{}
\mciteSetBstSublistMode{f}
\mciteSetBstMaxWidthForm{subitem}{(\alph{mcitesubitemcount})}
\mciteSetBstSublistLabelBeginEnd{\mcitemaxwidthsubitemform\space}
{\relax}{\relax}

\bibitem[Boyd(2008)]{B08_3}
Boyd,~R.~W. \emph{Nonlinear optics}, 3rd ed.;
\newblock Academic Press: Amsterdam, 2008\relax
\mciteBstWouldAddEndPuncttrue
\mciteSetBstMidEndSepPunct{\mcitedefaultmidpunct}
{\mcitedefaultendpunct}{\mcitedefaultseppunct}\relax
\EndOfBibitem
\bibitem[Garmire(2013)]{G13}
Garmire,~E. \emph{Opt.\ Express} \textbf{2013}, \emph{21}, 30532--30544\relax
\mciteBstWouldAddEndPuncttrue
\mciteSetBstMidEndSepPunct{\mcitedefaultmidpunct}
{\mcitedefaultendpunct}{\mcitedefaultseppunct}\relax
\EndOfBibitem
\bibitem[Kauranen and Zayats(2012)]{KZ12}
Kauranen,~M.; Zayats,~A.~V. \emph{Nat. Photon.} \textbf{2012}, \emph{6},
  737--748\relax
\mciteBstWouldAddEndPuncttrue
\mciteSetBstMidEndSepPunct{\mcitedefaultmidpunct}
{\mcitedefaultendpunct}{\mcitedefaultseppunct}\relax
\EndOfBibitem
\bibitem[Khurgin and Sun(2013)]{KS13}
Khurgin,~J.~B.; Sun,~G. \emph{Opt.\ Express} \textbf{2013}, \emph{21},
  27460--27480\relax
\mciteBstWouldAddEndPuncttrue
\mciteSetBstMidEndSepPunct{\mcitedefaultmidpunct}
{\mcitedefaultendpunct}{\mcitedefaultseppunct}\relax
\EndOfBibitem
\bibitem[Stockman(2011)]{S11}
Stockman,~M.~I. \emph{Phys.\ Today} \textbf{2011}, \emph{February},
  39--44\relax
\mciteBstWouldAddEndPuncttrue
\mciteSetBstMidEndSepPunct{\mcitedefaultmidpunct}
{\mcitedefaultendpunct}{\mcitedefaultseppunct}\relax
\EndOfBibitem
\bibitem[Atwater and Polman(2010)]{AP10}
Atwater,~H.~A.; Polman,~A. \emph{Nat.\ Mater.} \textbf{2010}, \emph{9},
  205--213\relax
\mciteBstWouldAddEndPuncttrue
\mciteSetBstMidEndSepPunct{\mcitedefaultmidpunct}
{\mcitedefaultendpunct}{\mcitedefaultseppunct}\relax
\EndOfBibitem
\bibitem[Vance et~al.(1997)Vance, Lemon, and Hupp]{VLH98}
Vance,~F.~W.; Lemon,~B.~I.; Hupp,~J.~T. \emph{J.\ Phys.\ Chem.\ B}
  \textbf{1997}, \emph{102}, 10091--10093\relax
\mciteBstWouldAddEndPuncttrue
\mciteSetBstMidEndSepPunct{\mcitedefaultmidpunct}
{\mcitedefaultendpunct}{\mcitedefaultseppunct}\relax
\EndOfBibitem
\bibitem[Russier-Antoine et~al.(2007)Russier-Antoine, Benichou, Bachelier,
  Jonin, and Brevet]{RBB07}
Russier-Antoine,~I.; Benichou,~E.; Bachelier,~G.; Jonin,~C.; Brevet,~P.~F.
  \emph{J.\ Phys.\ Chem.\ C} \textbf{2007}, \emph{111}, 9044--9048\relax
\mciteBstWouldAddEndPuncttrue
\mciteSetBstMidEndSepPunct{\mcitedefaultmidpunct}
{\mcitedefaultendpunct}{\mcitedefaultseppunct}\relax
\EndOfBibitem
\bibitem[Bao and Loh(2012)]{BL12}
Bao,~Q.; Loh,~K.~P. \emph{ACS\ Nano} \textbf{2012}, \emph{6}, 3677--3694\relax
\mciteBstWouldAddEndPuncttrue
\mciteSetBstMidEndSepPunct{\mcitedefaultmidpunct}
{\mcitedefaultendpunct}{\mcitedefaultseppunct}\relax
\EndOfBibitem
\bibitem[Koppens et~al.(2011)Koppens, Chang, and {Garc\'{\i}a de
  Abajo}]{paper176}
Koppens,~F. H.~L.; Chang,~D.~E.; {Garc\'{\i}a de Abajo},~F.~J. \emph{Nano\
  Lett.} \textbf{2011}, \emph{11}, 3370--3377\relax
\mciteBstWouldAddEndPuncttrue
\mciteSetBstMidEndSepPunct{\mcitedefaultmidpunct}
{\mcitedefaultendpunct}{\mcitedefaultseppunct}\relax
\EndOfBibitem
\bibitem[Grigorenko et~al.(2012)Grigorenko, Polini, and Novoselov]{GPN12}
Grigorenko,~A.~N.; Polini,~M.; Novoselov,~K.~S. \emph{Nat.\ Photon.}
  \textbf{2012}, \emph{6}, 749--758\relax
\mciteBstWouldAddEndPuncttrue
\mciteSetBstMidEndSepPunct{\mcitedefaultmidpunct}
{\mcitedefaultendpunct}{\mcitedefaultseppunct}\relax
\EndOfBibitem
\bibitem[{Garc\'{\i}a de Abajo}(2014)]{paper235}
{Garc\'{\i}a de Abajo},~F.~J. \emph{ACS\ Photon.} \textbf{2014}, \emph{1},
  135--152\relax
\mciteBstWouldAddEndPuncttrue
\mciteSetBstMidEndSepPunct{\mcitedefaultmidpunct}
{\mcitedefaultendpunct}{\mcitedefaultseppunct}\relax
\EndOfBibitem
\bibitem[Ju et~al.(2011)Ju, Geng, Horng, Girit, Martin, Hao, Bechtel, Liang,
  Zettl, Shen, and Wang]{JGH11}
Ju,~L.; Geng,~B.; Horng,~J.; Girit,~C.; Martin,~M.; Hao,~Z.; Bechtel,~H.~A.;
  Liang,~X.; Zettl,~A.; Shen,~Y.~R.; Wang,~F. \emph{Nat.\ Nanotech.}
  \textbf{2011}, \emph{6}, 630--634\relax
\mciteBstWouldAddEndPuncttrue
\mciteSetBstMidEndSepPunct{\mcitedefaultmidpunct}
{\mcitedefaultendpunct}{\mcitedefaultseppunct}\relax
\EndOfBibitem
\bibitem[Yan et~al.(2012)Yan, Li, Li, Zhu, Avouris, and Xia]{YLL12}
Yan,~H.; Li,~Z.; Li,~X.; Zhu,~W.; Avouris,~P.; Xia,~F. \emph{Nano\ Lett.}
  \textbf{2012}, \emph{12}, 3766--3771\relax
\mciteBstWouldAddEndPuncttrue
\mciteSetBstMidEndSepPunct{\mcitedefaultmidpunct}
{\mcitedefaultendpunct}{\mcitedefaultseppunct}\relax
\EndOfBibitem
\bibitem[Fei et~al.(2011)Fei, Andreev, Bao, Zhang, McLeod, Wang, Stewart, Zhao,
  Dominguez, Thiemens, Fogler, Tauber, Castro-Neto, Lau, Keilmann, and
  Basov]{FAB11}
Fei,~Z. et~al. \emph{Nano\ Lett.} \textbf{2011}, \emph{11}, 4701--4705\relax
\mciteBstWouldAddEndPuncttrue
\mciteSetBstMidEndSepPunct{\mcitedefaultmidpunct}
{\mcitedefaultendpunct}{\mcitedefaultseppunct}\relax
\EndOfBibitem
\bibitem[Yan et~al.(2012)Yan, Li, Chandra, Tulevski, Wu, Freitag, Zhu, Avouris,
  and Xia]{YLC12}
Yan,~H.; Li,~X.; Chandra,~B.; Tulevski,~G.; Wu,~Y.; Freitag,~M.; Zhu,~W.;
  Avouris,~P.; Xia,~F. \emph{Nat.\ Nanotech.} \textbf{2012}, \emph{7},
  330--334\relax
\mciteBstWouldAddEndPuncttrue
\mciteSetBstMidEndSepPunct{\mcitedefaultmidpunct}
{\mcitedefaultendpunct}{\mcitedefaultseppunct}\relax
\EndOfBibitem
\bibitem[Chen et~al.(2012)Chen, Badioli, Alonso-Gonz\'alez, Thongrattanasiri,
  Huth, Osmond, Spasenovi\'c, Centeno, Pesquera, Godignon, {Zurutuza Elorza},
  Camara, {Garc\'{\i}a de Abajo}, Hillenbrand, and Koppens]{paper196}
Chen,~J.; Badioli,~M.; Alonso-Gonz\'alez,~P.; Thongrattanasiri,~S.; Huth,~F.;
  Osmond,~J.; Spasenovi\'c,~M.; Centeno,~A.; Pesquera,~A.; Godignon,~P.;
  {Zurutuza Elorza},~A.; Camara,~N.; {Garc\'{\i}a de Abajo},~F.~J.;
  Hillenbrand,~R.; Koppens,~F. H.~L. \emph{Nature} \textbf{2012}, \emph{487},
  77--81\relax
\mciteBstWouldAddEndPuncttrue
\mciteSetBstMidEndSepPunct{\mcitedefaultmidpunct}
{\mcitedefaultendpunct}{\mcitedefaultseppunct}\relax
\EndOfBibitem
\bibitem[Fei et~al.(2012)Fei, Rodin, Andreev, Bao, McLeod, Wagner, Zhang, Zhao,
  Thiemens, Dominguez, Fogler, Neto, Lau, Keilmann, and Basov]{FRA12}
Fei,~Z.; Rodin,~A.~S.; Andreev,~G.~O.; Bao,~W.; McLeod,~A.~S.; Wagner,~M.;
  Zhang,~L.~M.; Zhao,~Z.; Thiemens,~M.; Dominguez,~G.; Fogler,~M.~M.; Neto,~A.
  H.~C.; Lau,~C.~N.; Keilmann,~F.; Basov,~D.~N. \emph{Nature} \textbf{2012},
  \emph{487}, 82--85\relax
\mciteBstWouldAddEndPuncttrue
\mciteSetBstMidEndSepPunct{\mcitedefaultmidpunct}
{\mcitedefaultendpunct}{\mcitedefaultseppunct}\relax
\EndOfBibitem
\bibitem[Mikhailov(2007)]{M07_2}
Mikhailov,~S.~A. \emph{Europhys.\ Lett.} \textbf{2007}, \emph{79}, 27002\relax
\mciteBstWouldAddEndPuncttrue
\mciteSetBstMidEndSepPunct{\mcitedefaultmidpunct}
{\mcitedefaultendpunct}{\mcitedefaultseppunct}\relax
\EndOfBibitem
\bibitem[Hendry et~al.(2010)Hendry, Hale, Moger, Savchenko, and
  Mikhailov]{HHM10}
Hendry,~E.; Hale,~P.~J.; Moger,~J.; Savchenko,~A.~K.; Mikhailov,~S.~A.
  \emph{Phys.\ Rev.\ Lett.} \textbf{2010}, \emph{105}, 097401\relax
\mciteBstWouldAddEndPuncttrue
\mciteSetBstMidEndSepPunct{\mcitedefaultmidpunct}
{\mcitedefaultendpunct}{\mcitedefaultseppunct}\relax
\EndOfBibitem
\bibitem[Zhang et~al.(2012)Zhang, Virally, Bao, Ping, Massar, Godbout, and
  Kockaert]{ZVB12}
Zhang,~H.; Virally,~S.; Bao,~Q.; Ping,~L.~K.; Massar,~S.; Godbout,~N.;
  Kockaert,~P. \emph{Opt.\ Lett.} \textbf{2012}, \emph{37}, 1856--1858\relax
\mciteBstWouldAddEndPuncttrue
\mciteSetBstMidEndSepPunct{\mcitedefaultmidpunct}
{\mcitedefaultendpunct}{\mcitedefaultseppunct}\relax
\EndOfBibitem
\bibitem[Kumar et~al.(2013)Kumar, Kumar, Gerstenkorn, Wang, Chiu, Smirl, and
  Zhao]{KKG13}
Kumar,~N.; Kumar,~J.; Gerstenkorn,~C.; Wang,~R.; Chiu,~H.-Y.; Smirl,~A.~L.;
  Zhao,~H. \emph{Phys.\ Rev.\ B} \textbf{2013}, \emph{87}, 121406(R)\relax
\mciteBstWouldAddEndPuncttrue
\mciteSetBstMidEndSepPunct{\mcitedefaultmidpunct}
{\mcitedefaultendpunct}{\mcitedefaultseppunct}\relax
\EndOfBibitem
\bibitem[Mikhailov(2011)]{M11}
Mikhailov,~S.~A. \emph{Phys.\ Rev.\ B} \textbf{2011}, \emph{84}, 045432\relax
\mciteBstWouldAddEndPuncttrue
\mciteSetBstMidEndSepPunct{\mcitedefaultmidpunct}
{\mcitedefaultendpunct}{\mcitedefaultseppunct}\relax
\EndOfBibitem
\bibitem[Gullans et~al.(2013)Gullans, Chang, Koppens, {Garc\'{\i}a de Abajo},
  and Lukin]{paper226}
Gullans,~M.; Chang,~D.~E.; Koppens,~F. H.~L.; {Garc\'{\i}a de Abajo},~F.~J.;
  Lukin,~M.~D. \emph{Phys.\ Rev.\ Lett.} \textbf{2013}, \emph{111},
  247401\relax
\mciteBstWouldAddEndPuncttrue
\mciteSetBstMidEndSepPunct{\mcitedefaultmidpunct}
{\mcitedefaultendpunct}{\mcitedefaultseppunct}\relax
\EndOfBibitem
\bibitem[MSG()]{MSG14}
 Manzoni, M. T.; Silveiro, I.; {Garc\'{\i}a de Abajo}, F. J.; Chang, D. E.
  Second-order quantum nonlinear optical processes in graphene nanostructures.
  {\it arXiv:1406.4360}.\relax
\mciteBstWouldAddEndPunctfalse
\mciteSetBstMidEndSepPunct{\mcitedefaultmidpunct}
{}{\mcitedefaultseppunct}\relax
\EndOfBibitem
\bibitem[Thongrattanasiri et~al.(2012)Thongrattanasiri, Manjavacas, and
  {Garc\'{\i}a de Abajo}]{paper183}
Thongrattanasiri,~S.; Manjavacas,~A.; {Garc\'{\i}a de Abajo},~F.~J. \emph{ACS\
  Nano} \textbf{2012}, \emph{6}, 1766--1775\relax
\mciteBstWouldAddEndPuncttrue
\mciteSetBstMidEndSepPunct{\mcitedefaultmidpunct}
{\mcitedefaultendpunct}{\mcitedefaultseppunct}\relax
\EndOfBibitem
\bibitem[Manjavacas et~al.(2013)Manjavacas, Marchesin, Thongrattanasiri, Koval,
  Nordlander, S\'{a}nchez-Portal, and {Garc\'{\i}a de Abajo}]{paper215}
Manjavacas,~A.; Marchesin,~F.; Thongrattanasiri,~S.; Koval,~P.; Nordlander,~P.;
  S\'{a}nchez-Portal,~D.; {Garc\'{\i}a de Abajo},~F.~J. \emph{ACS Nano}
  \textbf{2013}, \emph{7}, 3635--3643\relax
\mciteBstWouldAddEndPuncttrue
\mciteSetBstMidEndSepPunct{\mcitedefaultmidpunct}
{\mcitedefaultendpunct}{\mcitedefaultseppunct}\relax
\EndOfBibitem
\bibitem[Manjavacas et~al.(2013)Manjavacas, Thongrattanasiri, and {Garc\'{\i}a
  de Abajo}]{paper214}
Manjavacas,~A.; Thongrattanasiri,~S.; {Garc\'{\i}a de Abajo},~F.~J.
  \emph{Nanophotonics} \textbf{2013}, \emph{2}, 139--151\relax
\mciteBstWouldAddEndPuncttrue
\mciteSetBstMidEndSepPunct{\mcitedefaultmidpunct}
{\mcitedefaultendpunct}{\mcitedefaultseppunct}\relax
\EndOfBibitem
\bibitem[Pines and Nozi\`{e}res(1966)]{PN1966}
Pines,~D.; Nozi\`{e}res,~P. \emph{The Theory of Quantum Liquids};
\newblock W. A. Benjamin, Inc.: New York, 1966\relax
\mciteBstWouldAddEndPuncttrue
\mciteSetBstMidEndSepPunct{\mcitedefaultmidpunct}
{\mcitedefaultendpunct}{\mcitedefaultseppunct}\relax
\EndOfBibitem
\bibitem[Galletto et~al.(1999)Galletto, Brevet, Girault, Antoine, and
  Broyer]{GBG99}
Galletto,~P.; Brevet,~P.~F.; Girault,~H.~H.; Antoine,~R.; Broyer,~M.
  \emph{Chem.\ Commun.} \textbf{1999},  581--582\relax
\mciteBstWouldAddEndPuncttrue
\mciteSetBstMidEndSepPunct{\mcitedefaultmidpunct}
{\mcitedefaultendpunct}{\mcitedefaultseppunct}\relax
\EndOfBibitem
\bibitem[Russier-Antoine et~al.(2010)Russier-Antoine, Duboisset, Bachelier,
  Benichou, Jonin, Fatti, Vall\'{e}e, S\'{a}nchez-Iglesias, Pastoriza-Santos,
  Liz-Marz\'an, and Brevet]{RDB10}
Russier-Antoine,~I.; Duboisset,~J.; Bachelier,~G.; Benichou,~E.; Jonin,~C.;
  Fatti,~N.~D.; Vall\'{e}e,~F.; S\'{a}nchez-Iglesias,~A.; Pastoriza-Santos,~I.;
  Liz-Marz\'an,~L.~M.; Brevet,~P.-F. \emph{J.\ Phys.\ Chem.\ Lett.}
  \textbf{2010}, \emph{1}, 874--880\relax
\mciteBstWouldAddEndPuncttrue
\mciteSetBstMidEndSepPunct{\mcitedefaultmidpunct}
{\mcitedefaultendpunct}{\mcitedefaultseppunct}\relax
\EndOfBibitem
\bibitem[Singh et~al.(2009)Singh, Senapati, Neely, Kolawole, Hawker, and
  Ray]{SSN09}
Singh,~A.~K.; Senapati,~D.; Neely,~A.; Kolawole,~G.; Hawker,~C.; Ray,~P.~C.
  \emph{Chem.\ Phys.\ Lett.} \textbf{2009}, \emph{481}, 94--98\relax
\mciteBstWouldAddEndPuncttrue
\mciteSetBstMidEndSepPunct{\mcitedefaultmidpunct}
{\mcitedefaultendpunct}{\mcitedefaultseppunct}\relax
\EndOfBibitem
\bibitem[Uchida et~al.(1994)Uchida, Kaneko, Omi, Hata, Tanji, Asahara,
  Ikushima, Tokizaki, and Nakamura]{UKO94}
Uchida,~K.; Kaneko,~S.; Omi,~S.; Hata,~C.; Tanji,~H.; Asahara,~Y.;
  Ikushima,~A.~J.; Tokizaki,~T.; Nakamura,~A. \emph{J.\ Opt.\ Soc.\ Am.\ B}
  \textbf{1994}, \emph{11}, 1236--1243\relax
\mciteBstWouldAddEndPuncttrue
\mciteSetBstMidEndSepPunct{\mcitedefaultmidpunct}
{\mcitedefaultendpunct}{\mcitedefaultseppunct}\relax
\EndOfBibitem
\bibitem[Liu et~al.(2006)Liu, Tai, Yu, Wen, Chu, Chen, Prasad, Lin, and
  Sun]{LTY06}
Liu,~T.-M.; Tai,~S.-P.; Yu,~C.-H.; Wen,~Y.-C.; Chu,~S.-W.; Chen,~L.-J.;
  Prasad,~M.~R.; Lin,~K.-J.; Sun,~C.-K. \emph{Appl.\ Phys.\ Lett.}
  \textbf{2006}, \emph{89}, 043122\relax
\mciteBstWouldAddEndPuncttrue
\mciteSetBstMidEndSepPunct{\mcitedefaultmidpunct}
{\mcitedefaultendpunct}{\mcitedefaultseppunct}\relax
\EndOfBibitem
\bibitem[Boyd et~al.(2014)Boyd, Shi, and {De Leon}]{BSD14}
Boyd,~R.~W.; Shi,~Z.; {De Leon},~I. \emph{Opt.\ Commun.} \textbf{2014},
  \emph{326}, 74--79\relax
\mciteBstWouldAddEndPuncttrue
\mciteSetBstMidEndSepPunct{\mcitedefaultmidpunct}
{\mcitedefaultendpunct}{\mcitedefaultseppunct}\relax
\EndOfBibitem
\bibitem[Luo et~al.(2011)Luo, Shiao, and Huang]{LYG11}
Luo,~Y.~L.; Shiao,~Y.~S.; Huang,~Y.~F. \emph{Nat.\ Nanotech.} \textbf{2011},
  \emph{6}, 247--252\relax
\mciteBstWouldAddEndPuncttrue
\mciteSetBstMidEndSepPunct{\mcitedefaultmidpunct}
{\mcitedefaultendpunct}{\mcitedefaultseppunct}\relax
\EndOfBibitem
\bibitem[Subramaniam et~al.(2012)Subramaniam, Libisch, Li, Pauly, Geringer,
  Reiter, Mashoff, Liebmann, Burgd\"orfer, Busse, Michely, Mazzarello, Pratzer,
  and Morgenstern]{SLL12}
Subramaniam,~D.; Libisch,~F.; Li,~Y.; Pauly,~C.; Geringer,~V.; Reiter,~R.;
  Mashoff,~T.; Liebmann,~M.; Burgd\"orfer,~J.; Busse,~C.; Michely,~T.;
  Mazzarello,~R.; Pratzer,~M.; Morgenstern,~M. \emph{Phys.\ Rev.\ Lett.}
  \textbf{2012}, \emph{108}, 046801\relax
\mciteBstWouldAddEndPuncttrue
\mciteSetBstMidEndSepPunct{\mcitedefaultmidpunct}
{\mcitedefaultendpunct}{\mcitedefaultseppunct}\relax
\EndOfBibitem
\bibitem[Kim et~al.(2012)Kim, Hwang, Kim, Shin, Shin, Kim, Yang, Park, H.,
  Choi, Ko, Sim, Sone, Choi, Bae, and Hong]{KHK12}
Kim,~S. et~al. \emph{ACS\ Nano} \textbf{2012}, \emph{6}, 8203--8208\relax
\mciteBstWouldAddEndPuncttrue
\mciteSetBstMidEndSepPunct{\mcitedefaultmidpunct}
{\mcitedefaultendpunct}{\mcitedefaultseppunct}\relax
\EndOfBibitem
\bibitem[Wu et~al.(2007)Wu, Pisula, and M\"ullen]{WPM07}
Wu,~J.; Pisula,~W.; M\"ullen,~K. \emph{Chem.\ Rev.} \textbf{2007}, \emph{107},
  718--747\relax
\mciteBstWouldAddEndPuncttrue
\mciteSetBstMidEndSepPunct{\mcitedefaultmidpunct}
{\mcitedefaultendpunct}{\mcitedefaultseppunct}\relax
\EndOfBibitem
\bibitem[Feng et~al.(2009)Feng, Pisula, and M\"ullen]{FPM09}
Feng,~X.; Pisula,~W.; M\"ullen,~K. \emph{Pure\ Appl.\ Chem.} \textbf{2009},
  \emph{81}, 2203--2224\relax
\mciteBstWouldAddEndPuncttrue
\mciteSetBstMidEndSepPunct{\mcitedefaultmidpunct}
{\mcitedefaultendpunct}{\mcitedefaultseppunct}\relax
\EndOfBibitem
\bibitem[Luo et~al.(2012)Luo, Liu, and Zhi]{BLZ12}
Luo,~B.; Liu,~S.; Zhi,~L. \emph{Small} \textbf{2012}, \emph{8}, 630--646\relax
\mciteBstWouldAddEndPuncttrue
\mciteSetBstMidEndSepPunct{\mcitedefaultmidpunct}
{\mcitedefaultendpunct}{\mcitedefaultseppunct}\relax
\EndOfBibitem
\bibitem[Fang et~al.(2013)Fang, Thongrattanasiri, Schlather, Liu, Ma, Wang,
  Ajayan, Nordlander, Halas, and {Garc\'{\i}a de Abajo}]{paper212}
Fang,~Z.; Thongrattanasiri,~S.; Schlather,~A.; Liu,~Z.; Ma,~L.; Wang,~Y.;
  Ajayan,~P.~M.; Nordlander,~P.; Halas,~N.~J.; {Garc\'{\i}a de Abajo},~F.~J.
  \emph{ACS Nano} \textbf{2013}, \emph{7}, 2388--2395\relax
\mciteBstWouldAddEndPuncttrue
\mciteSetBstMidEndSepPunct{\mcitedefaultmidpunct}
{\mcitedefaultendpunct}{\mcitedefaultseppunct}\relax
\EndOfBibitem
\bibitem[Wallace(1947)]{W1947}
Wallace,~P.~R. \emph{Phys.\ Rev.} \textbf{1947}, \emph{71}, 622--634\relax
\mciteBstWouldAddEndPuncttrue
\mciteSetBstMidEndSepPunct{\mcitedefaultmidpunct}
{\mcitedefaultendpunct}{\mcitedefaultseppunct}\relax
\EndOfBibitem
\bibitem[{Castro Neto} et~al.(2009){Castro Neto}, Guinea, Peres, Novoselov, and
  Geim]{CGP09}
{Castro Neto},~A.~H.; Guinea,~F.; Peres,~N. M.~R.; Novoselov,~K.~S.;
  Geim,~A.~K. \emph{Rev.\ Mod.\ Phys.} \textbf{2009}, \emph{81}, 109--162\relax
\mciteBstWouldAddEndPuncttrue
\mciteSetBstMidEndSepPunct{\mcitedefaultmidpunct}
{\mcitedefaultendpunct}{\mcitedefaultseppunct}\relax
\EndOfBibitem
\bibitem[Hedin and Lundqvist(1970)]{HL1970}
Hedin,~L.; Lundqvist,~S. Effects of Electron-Electron and Electron-Phonon
  Interactions on the One-Electron States of Solids. In \emph{Solid State
  Physics}; Frederick~Seitz,~D.~T., Ehrenreich,~H., Eds.;
\newblock Academic Press, 1970;
\newblock Vol.~23, pp 1 -- 181\relax
\mciteBstWouldAddEndPuncttrue
\mciteSetBstMidEndSepPunct{\mcitedefaultmidpunct}
{\mcitedefaultendpunct}{\mcitedefaultseppunct}\relax
\EndOfBibitem
\bibitem[Mermin(1970)]{M1970}
Mermin,~N.~D. \emph{Phys.\ Rev.\ B} \textbf{1970}, \emph{1}, 2362--2363\relax
\mciteBstWouldAddEndPuncttrue
\mciteSetBstMidEndSepPunct{\mcitedefaultmidpunct}
{\mcitedefaultendpunct}{\mcitedefaultseppunct}\relax
\EndOfBibitem
\bibitem[Potasz et~al.(2012)Potasz, G\"u\ifmmode~\mbox{\c{c}}\else
  \c{c}\fi{}l\"u, W\'ojs, and Hawrylak]{PGW12}
Potasz,~P.; G\"u\ifmmode~\mbox{\c{c}}\else \c{c}\fi{}l\"u,~A.~D.; W\'ojs,~A.;
  Hawrylak,~P. \emph{Phys.\ Rev.\ B} \textbf{2012}, \emph{85}, 075431\relax
\mciteBstWouldAddEndPuncttrue
\mciteSetBstMidEndSepPunct{\mcitedefaultmidpunct}
{\mcitedefaultendpunct}{\mcitedefaultseppunct}\relax
\EndOfBibitem
\bibitem[Kanis et~al.(1994)Kanis, Ratner, and Marks]{KRM94}
Kanis,~D.~R.; Ratner,~M.~A.; Marks,~T.~J. \emph{Chem. Reviews} \textbf{1994},
  \emph{94}, 195--242\relax
\mciteBstWouldAddEndPuncttrue
\mciteSetBstMidEndSepPunct{\mcitedefaultmidpunct}
{\mcitedefaultendpunct}{\mcitedefaultseppunct}\relax
\EndOfBibitem
\bibitem[Flytzanis and Tang(1980)]{FT1980}
Flytzanis,~C.; Tang,~C.~L. \emph{Phys.\ Rev.\ Lett.} \textbf{1980}, \emph{45},
  441--445\relax
\mciteBstWouldAddEndPuncttrue
\mciteSetBstMidEndSepPunct{\mcitedefaultmidpunct}
{\mcitedefaultendpunct}{\mcitedefaultseppunct}\relax
\EndOfBibitem
\bibitem[Ritchie and Bowden(1985)]{RB1985}
Ritchie,~B.; Bowden,~C.~M. \emph{Phys.\ Rev.\ A} \textbf{1985}, \emph{32},
  2293--2297\relax
\mciteBstWouldAddEndPuncttrue
\mciteSetBstMidEndSepPunct{\mcitedefaultmidpunct}
{\mcitedefaultendpunct}{\mcitedefaultseppunct}\relax
\EndOfBibitem
\bibitem[Novoselov et~al.(2004)Novoselov, Geim, Morozov, Jiang, Zhang, Dubonos,
  Grigorieva, and Firsov]{NGM04}
Novoselov,~K.~S.; Geim,~A.~K.; Morozov,~S.~V.; Jiang,~D.; Zhang,~Y.;
  Dubonos,~S.~V.; Grigorieva,~I.~V.; Firsov,~A.~A. \emph{Science}
  \textbf{2004}, \emph{306}, 666--669\relax
\mciteBstWouldAddEndPuncttrue
\mciteSetBstMidEndSepPunct{\mcitedefaultmidpunct}
{\mcitedefaultendpunct}{\mcitedefaultseppunct}\relax
\EndOfBibitem
\bibitem[Novoselov et~al.(2005)Novoselov, Geim, Morozov, Jiang, Katsnelson,
  Grigorieva, Dubonos, and Firsov]{NGM05}
Novoselov,~K.~S.; Geim,~A.~K.; Morozov,~S.~V.; Jiang,~D.; Katsnelson,~M.~I.;
  Grigorieva,~I.~V.; Dubonos,~S.~V.; Firsov,~A.~A. \emph{Nature} \textbf{2005},
  \emph{438}, 197--200\relax
\mciteBstWouldAddEndPuncttrue
\mciteSetBstMidEndSepPunct{\mcitedefaultmidpunct}
{\mcitedefaultendpunct}{\mcitedefaultseppunct}\relax
\EndOfBibitem
\bibitem[Jablan et~al.(2009)Jablan, Buljan, and {Solja\v{c}i\'{c}}]{JBS09}
Jablan,~M.; Buljan,~H.; {Solja\v{c}i\'{c}},~M. \emph{Phys.\ Rev.\ B}
  \textbf{2009}, \emph{80}, 245435\relax
\mciteBstWouldAddEndPuncttrue
\mciteSetBstMidEndSepPunct{\mcitedefaultmidpunct}
{\mcitedefaultendpunct}{\mcitedefaultseppunct}\relax
\EndOfBibitem
\bibitem[Thongrattanasiri et~al.(2012)Thongrattanasiri, Koppens, and
  {Garc\'{\i}a de Abajo}]{paper182}
Thongrattanasiri,~S.; Koppens,~F. H.~L.; {Garc\'{\i}a de Abajo},~F.~J.
  \emph{Phys.\ Rev.\ Lett.} \textbf{2012}, \emph{108}, 047401\relax
\mciteBstWouldAddEndPuncttrue
\mciteSetBstMidEndSepPunct{\mcitedefaultmidpunct}
{\mcitedefaultendpunct}{\mcitedefaultseppunct}\relax
\EndOfBibitem
\end{mcitethebibliography}

\providecommand*{\mcitethebibliography}{\thebibliography}
\csname @ifundefined\endcsname{endmcitethebibliography}
{\let\endmcitethebibliography\endthebibliography}{}

\end{document}